\documentclass[twoside,leqno,twocolumn]{article}

\usepackage[letterpaper]{geometry}

\usepackage{siamproceedings}

\usepackage[T1]{fontenc}
\usepackage{amsfonts}
\usepackage{bm}
\usepackage{booktabs}
\usepackage{graphicx}
\graphicspath{{images/}}
\usepackage{subcaption}
\usepackage{algorithmic}
\usepackage{float}
\restylefloat{algorithm}
\usepackage{pgfplots}
\pgfplotsset{compat=1.18}
\usepackage{tikz}
\usetikzlibrary{arrows.meta,positioning,calc,shapes.geometric,fit,backgrounds}
\usepackage{multirow}
\usepackage{enumitem}
\usepackage{placeins}

\definecolor{cblue}{HTML}{0072B2}
\definecolor{corange}{HTML}{D55E00}
\definecolor{cgreen}{HTML}{009E73}
\definecolor{cred}{HTML}{CC79A7}
\definecolor{cgray}{HTML}{999999}
\definecolor{bandgray}{HTML}{EDF2F7}
\usepackage{colortbl}

\newcommand{\vect}[1]{\bm{#1}}
\newcommand{\vu}{\vect{u}}
\newcommand{\vv}{\vect{v}}
\newcommand{\vk}{\vect{k}}
\newcommand{\vz}{\vect{z}}
\newcommand{\veps}{\vect{\epsilon}}
\newcommand{\R}{\mathbb{R}}
\newcommand{\E}{\mathbb{E}}


\begin{document}

\newcommand\relatedversion{}

\title{\Large FlowRefiner: Flow Matching-Based Iterative Refinement\\for 3D Turbulent Flow Simulation\relatedversion}
\author{
\begin{tabular}{ccc}
Yilong Dai\thanks{University of Alabama, \texttt{ydai17@ua.edu}} \qquad \qquad &
Yiming Sun\thanks{University of Pittsburgh, \texttt{yimingsun@pitt.edu}} \qquad \qquad &
Yiheng Chen\thanks{University of Alabama, \texttt{ychen226@ua.edu}}
\end{tabular}
\\[0.3em]
\begin{tabular}{ccc}
Shengyu Chen\thanks{University of Pittsburgh, \texttt{shc160@pitt.edu}} \qquad \qquad &
Xiaowei Jia\thanks{University of Pittsburgh, \texttt{xiaowei@pitt.edu}} \qquad \qquad &
Runlong Yu\thanks{University of Alabama, \texttt{ryu5@ua.edu}. Runlong Yu is the corresponding author.}
\end{tabular}
}
\date{}

\maketitle


\begin{abstract}
Accurate autoregressive prediction of 3D turbulent flows remains challenging for neural PDE solvers, as small errors in fine-scale structures can accumulate rapidly over rollout. In this paper, we propose \textbf{FlowRefiner}, a flow matching-based iterative refinement framework for 3D turbulent flow simulation. The method replaces stochastic denoising refinement with deterministic ODE-based correction, uses a unified velocity-field regression objective across all refinement stages, and introduces a decoupled sigma schedule that fixes the noise range independently of refinement depth. These design choices yield stable and effective refinement in the small-noise regime. Experiments on large-scale 3D turbulence with rich multi-scale structures show that FlowRefiner achieves state-of-the-art autoregressive prediction accuracy and strong physical consistency. Although developed for turbulent flow simulation, the proposed framework is broadly applicable to iterative refinement problems in scientific modeling.
\end{abstract}

\section{Introduction.}\label{sec:intro}

Turbulent flow simulation is central to science and engineering, yet high-fidelity three-dimensional computation remains expensive. Direct numerical simulation (DNS) resolves the full cascade of scales at prohibitive cost, with complexity that grows rapidly with Reynolds number, making long-horizon simulation impractical for many real settings. Data-driven surrogates therefore offer an attractive alternative: neural operators, U-Net-style architectures, and transformer-based PDE models can accelerate prediction by orders of magnitude while learning complex spatiotemporal dynamics directly from data~\cite{li2021fno,gupta2022modernunet,li2023oformer,wu2024transolver}. For these methods to be useful in practice, they should support stable and accurate autoregressive prediction of turbulent flows over extended horizons.

However, this remains challenging because turbulent flow prediction is not a standard next-step regression problem. Autoregressive neural PDE solvers are known to suffer from spectral bias: low-frequency components are learned more readily than high-frequency ones~\cite{rahaman2019spectral}, and standard $\ell_2$ training further emphasizes energetic large-scale modes over weaker but dynamically important fine structures~\cite{lippe2024pderefiner,dai2026learningpdesolversphysics}. As a result, a model may achieve reasonable one-step accuracy while still under-resolving sharp interfaces, filamentary structures, and other small-scale features whose errors compound rapidly when predictions are recursively fed back as inputs. Physics-informed objectives can partially mitigate this behavior by penalizing PDE residuals or enforcing constraints~\cite{raissi2019pinn,wang2021gradient,wang2024causal,dai2025pest,brandstetter2022message}, but they do not eliminate the central difficulty of repeated self-correction in a deterministic, multiscale dynamical system.

This has motivated a promising direction: iterative refinement. Instead of requiring a model to recover the next state in a single pass, refinement-based approaches first produce a coarse forecast and then improve it through a sequence of corrective steps. PDE-Refiner showed that diffusion-style denoising refinement can substantially improve learned PDE solvers on lower-dimensional benchmarks by forcing the model to revisit progressively finer details~\cite{lippe2024pderefiner}. This is an important conceptual shift. It reframes forecasting not as a one-shot prediction, but as predict-then-correct, which is naturally appealing for turbulent flows whose errors are concentrated in under-resolved fine scales.

Despite their promise, existing refinement methods have several limitations. Most inherit the DDPM formulation~\cite{ho2020ddpm}, which was designed for generative modeling from noise rather than iterative correction of an already informative prediction. In 3D turbulence, this mismatch becomes structural. First, stochastic denoising injects fresh variance at every refinement stage, even though the underlying Navier--Stokes dynamics are deterministic. Second, the learning objective changes semantics across stages: the base predictor regresses toward the physical solution, whereas refinement stages regress toward injected noise, creating a discontinuity in what the shared network is asked to learn. Third, the standard noise schedule couples perturbation magnitude to refinement depth, so increasing the number of refinement steps can also make the correction problem more difficult. In complex 3D flow fields, these effects are no longer minor implementation details but central obstacles to stable and effective refinement.

These limitations indicate that effective refinement requires a deterministic correction mechanism, a semantically consistent regression target across stages, and a refinement process whose depth can increase without enlarging the perturbation range. To fill this gap, we propose \textbf{FlowRefiner}, a flow matching (FM)-based iterative refinement framework for 3D turbulent flow prediction~\cite{lipman2023flow,liu2023rectified}. FlowRefiner replaces stochastic denoising with deterministic ODE-based correction, uses a unified velocity-field regression objective across all refinement stages, and introduces a decoupled sigma schedule that fixes the noise range independently of the number of refinement steps. Conceptually, FlowRefiner treats refinement as deterministic transport in state space from an initial forecast toward the target flow, rather than repeated denoising of newly injected randomness. An additional benefit is that FM lifts the Gaussian noise restriction and permits spectrally structured priors tailored to turbulence statistics~\cite{kingma2024diffusionmeetsflow}.

We evaluate FlowRefiner on two 3D turbulence benchmarks, forced isotropic turbulence (FIT) from JHTDB~\cite{li2008jhtdb,perlman2007jhtdb} and Taylor--Green vortex (TGV). On large-scale 3D isotropic turbulence with complex high-frequency structures, FlowRefiner achieves the best autoregressive accuracy among deterministic, physics-informed, and generative baselines while maintaining strong physical consistency.
Our contributions are summarized as follows:
\begin{itemize}[nosep,leftmargin=*]
\item We identify a diffusion--refinement mismatch for 3D turbulence: DDPM-style iterative correction introduces stochastic variance accumulation, regression-target discontinuity, and noise--depth coupling that limit effective refinement on complex flows.

\item We propose FlowRefiner, an FM-based iterative refinement framework that combines deterministic ODE-based correction, a unified velocity-field regression objective, and a decoupled sigma schedule for stable small-noise refinement.

\item We establish state-of-the-art autoregressive prediction on large-scale 3D isotropic turbulence with complex high-frequency structures, show strong physical consistency without explicit physics supervision, and clarify two key design findings: noise type is secondary, whereas physical projection under FM induces a fundamental accuracy--consistency trade-off.
\end{itemize}

\section{Related Work.}
\label{sec:related}

Neural PDE solvers have substantially improved data-driven forecasting of complex dynamics. Spectral operators, physical-space encoder--decoder models, and attention-based architectures have shown that high-dimensional spatiotemporal systems can be predicted efficiently without running full numerical solvers at inference time~\cite{li2021fno,tran2023ffno,gupta2022modernunet,li2023oformer,wu2024transolver,takamoto2022pdebench,wen2022u,hao2024dpot,li2023scalable}. A complementary line of work has incorporated governing equations into training through residual penalties or more structured physics-guided objectives~\cite{raissi2019pinn,wang2021gradient,wang2024causal,li2024physics,dai2025pest}. These advances have strengthened learned PDE forecasting but remain focused on improving one-shot prediction. For long autoregressive rollouts, the central unresolved difficulty is how to repeatedly correct small but dynamically important errors once they have entered the prediction trajectory.

This limitation has motivated iterative refinement. Rather than requiring a single forward pass to recover the full next state, refinement-based methods first produce an initial forecast and then improve it through a sequence of corrective updates. Existing work has shown that diffusion-based denoising can serve this role effectively, with PDE-Refiner providing the clearest formulation of post-hoc iterative correction for neural PDE solvers~\cite{lippe2024pderefiner,pdespectralrefiner2025,ho2020ddpm,shysheya2024conditional,yoo2026diffusionrollout}. More broadly, recent generative models for scientific dynamics have also explored flow-based transport formulations, including flow matching, which replaces stochastic denoising with deterministic ODE evolution and a simpler regression objective~\cite{lipman2023flow,liu2023rectified,albergo2023stochastic,kingma2024diffusionmeetsflow,song2021scorebased,dai2026flowlearnerspdesphysicstophysics}. The key gap, however, is not the lack of generative models for PDEs, but the lack of a refinement formulation designed for stable, repeated, small-noise correction of deterministic forecasts.

This gap is especially visible in the current separation between diffusion-style refinement and flow-matching-based PDE generation. Existing diffusion refiners inherit the variance accumulation, target discontinuity, and noise-schedule coupling of denoising formulations, while existing FM-based PDE methods have mainly focused on direct generation from noise or low-fidelity inputs rather than iterative correction around an already informative prediction~\cite{lienen2024from,kohl2024acdm,du2024confild,cfo2025,fourierflow2025,wu2025latentfm,residualfm2024,oommen2024integrating}. Our contribution lies in connecting these two lines: formulating flow matching as an iterative refinement mechanism for long-horizon turbulent prediction.

\section{Problem Formulation.}\label{sec:problem}

We consider autoregressive prediction for three-dimensional incompressible flow governed by the Navier--Stokes equations
\begin{equation}\label{eq:ns}
  \frac{\partial \vu}{\partial t} + (\vu \cdot \nabla)\vu
  = -\nabla p + \nu \nabla^2 \vu, \qquad
  \nabla \cdot \vu = 0,
\end{equation}
where $\vu(\vect{x},t) \in \R^3$ is the velocity field, $p(\vect{x},t) \in \R$ is the pressure field, and $\nu$ is the kinematic viscosity.

At each time step, we represent the flow state as
$
\mathbf{s}_t = [u_t, v_t, w_t, p_t],
$
where $u_t$, $v_t$, and $w_t$ denote the three velocity components and $p_t$ denotes pressure.
Given a block of $T_{\mathrm{in}}$ past states
$
\mathbf{S}_t = \{\mathbf{s}_{t-T_{\mathrm{in}}+1}, \ldots, \mathbf{s}_t\},
$
the learning task is to predict the next block of $T_{\mathrm{out}}$ states
$
\mathbf{Y}_t = \{\mathbf{s}_{t+1}, \ldots, \mathbf{s}_{t+T_{\mathrm{out}}}\}.
$
Let $f_\theta$ denote a learned predictor that maps an input block to a predicted output block,
$
\hat{\mathbf{Y}}_t = f_\theta(\mathbf{S}_t).
$
For long-horizon forecasting, prediction is performed autoregressively at the block level: the predicted block $\hat{\mathbf{Y}}_t$ is appended to the context and used to form the next input block, which is then fed back to predict the subsequent $T_{\mathrm{out}}$ states.
Repeating this for $R$ rounds yields $R \cdot T_{\mathrm{out}}$ predicted states
$
\hat{\mathbf{s}}_{t+1}, \hat{\mathbf{s}}_{t+2}, \ldots, \hat{\mathbf{s}}_{t + R \cdot T_{\mathrm{out}}}.
$
In our main experiments, $T_{\mathrm{in}} = T_{\mathrm{out}} = 5$.
Our goal is to improve the accuracy and stability of this autoregressive block prediction process.

\begin{figure*}[!t]
\centering
\includegraphics[width=\textwidth]{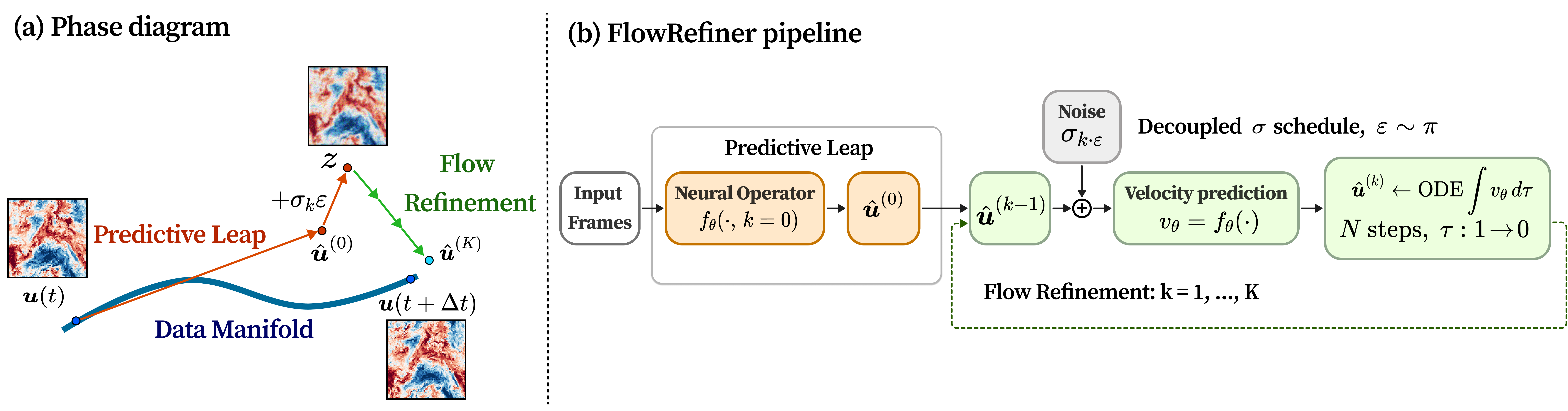}
\caption{FlowRefiner overview. \textbf{(a)} Predictive leap followed by flow refinement in state space. \textbf{(b)} Pipeline: the base model ($k{=}0$) produces an initial prediction; $K$ refinement steps inject noise and integrate the FM velocity field with a decoupled sigma schedule.}
\label{fig:pipeline}
\end{figure*}

\section{Method.}\label{sec:method} Here, we present FlowRefiner, a new FM-based iterative refinement framework for 3D turbulent flow prediction (\cref{fig:pipeline}). We first explain why denoising refinement becomes unstable, and then present the proposed refinement formulation, decoupled sigma schedule, and related design choices.

\subsection{Why Denoising Refinement Fails.}\label{sec:limitations}

Iterative refinement offers a natural alternative to one-shot prediction for turbulent flow forecasting.
Among existing approaches, PDE-Refiner~\cite{lippe2024pderefiner} provides a representative DDPM-based formulation of post-hoc refinement for neural PDE solvers~\cite{ho2020ddpm}.
It uses a single step-conditioned network for both base prediction and refinement: at $k{=}0$, the model predicts the clean solution from the input context, while at $k{\geq}1$, Gaussian noise $\sigma_k \veps_k$ is added to the ground-truth solution and the network is trained with a v-prediction objective~\cite{salimans2022progressive}, which in the small-noise regime is approximately equivalent to predicting the injected noise.
At inference time, the method alternates between perturbing the current estimate and denoising it in a single step.
Although effective on lower-dimensional benchmarks, it becomes problematic in 3D turbulence due to three limitations.

\noindent\textbf{(1) Noise--depth coupling.}
The schedule $\sigma_k = \sigma_{\min}^{k/K}$ ties the first-step noise level $\sigma_1 = \sigma_{\min}^{1/K}$ to the total number of refinement steps $K$.
Increasing refinement depth also increases the perturbation magnitude encountered at early steps.
On simple 1D or 2D systems, this coupling can be tolerable, but in 3D turbulence, the same fractional perturbation corrupts far more degrees of freedom and disproportionately destroys weak high-frequency structures. As a result, deeper refinement becomes even harder.

\noindent\textbf{(2) Regression-target discontinuity.}
The shared network is asked to solve two semantically different tasks.
At $k{=}0$, it regresses toward the physical solution; at $k{\geq}1$, it regresses toward injected noise.
These objectives impose different gradient directions on the same model, creating a mismatch between predicting the state itself and predicting the noise around that state.
In our experiments, this manifests as substantial training instability and loss oscillation.

\noindent\textbf{(3) Stochastic perturbation accumulation.}
The incompressible Navier--Stokes dynamics define a deterministic Markov process, but DDPM-style refinement injects fresh stochastic perturbations at every step of the autoregressive chain.
Because rollout has no mechanism for retrospective correction, these perturbations accumulate forward in time.
This is especially undesirable when refinement is meant to correct a nearly correct deterministic forecast rather than sample from a multimodal data distribution.

\subsection{FM-Based Iterative Refinement.}\label{sec:flowrefiner}

To resolve the above mismatch, we reformulate iterative refinement through flow matching (FM)~\cite{lipman2023flow,liu2023rectified}.
FM defines transport between a data point and a noise endpoint through linear interpolation:
\begin{equation}\label{eq:fm_interp}
  \vz_\tau = (1-\tau)\,\vect{x}_1 + \tau\,\veps, \qquad \tau \in [0,1],
\end{equation}
where $\vect{x}_1$ is a clean sample and $\veps$ is drawn from a noise prior.
The corresponding velocity field is
$
  \vv_\tau = \veps - \vect{x}_1,
$
and learning fits a model $\vv_\theta(\vz_\tau,\tau)$ to this target.
Inference then solves the deterministic ODE:
\begin{equation}
  \frac{d\vz}{d\tau} = \vv_\theta(\vz_\tau,\tau)
\end{equation}
from $\tau{=}1$ to $\tau{=}0$.
Unlike DDPM, FM imposes no Gaussian restriction on $\veps$.

FlowRefiner uses FM not for direct generation from noise, but for local refinement around a base forecast.
The key idea is to interpret each refinement stage as deterministic transport from a perturbed local state back toward the clean target.
This makes every refinement step solve the same semantic task: move the current estimate toward the ground truth.
In contrast to denoising refinement, where perfect noise prediction can yield no net correction, FM makes the correction direction itself the learning target.
A single model $f_\theta$ serves two roles.
At step $k{=}0$, it produces a deterministic base prediction.
At steps $k{\geq}1$, it predicts the FM velocity field used for refinement.

\noindent \textbf{Training.}
Given a training pair $(\vu_{\mathrm{in}}, \vu_{\mathrm{gt}})$, we uniformly sample a refinement step
$k \sim \{0,1,\ldots,K\}$.
For $k{=}0$, we use standard MSE regression:
\begin{equation}\label{eq:loss_k0}
  \mathcal{L}_0 = \|f_\theta(\vu_{\mathrm{in}}, k{=}0) - \vu_{\mathrm{gt}}\|^2.
\end{equation}
For $k{\geq}1$, we uniformly sample $\tau \sim [0,1]$ and $\veps \sim \pi$, where $\pi$ is the chosen noise prior, and define a localized FM interpolation as:
\begin{equation}
  \vz_\tau = (1-\tau)\,\vu_{\mathrm{gt}} + \tau\,\sigma_k\,\veps.
\end{equation}
This interpolation spans only a small neighborhood around the ground truth.
The target velocity is the normalized direction from the noisy state toward the clean data, given as:
\begin{equation}\label{eq:loss_kge1}
  \mathcal{L}_k
  = \|f_\theta(\vu_{\mathrm{in}}, \vz_\tau, k, \tau) - (-\veps)\|^2.
\end{equation}
The overall objective is
$
  \mathcal{L} = \E_k[\mathcal{L}_k].
$

\noindent \textbf{Inference.}
Starting from the base prediction
$\hat{\vu} = f_\theta(\vu_{\mathrm{in}}, k{=}0)$,
we apply $K$ refinement steps.
At each step $k$, we perturb the current estimate by $\sigma_k \veps$ and then integrate the learned velocity field from $\tau{=}1$ to $\tau{=}0$ using $N$ Euler substeps:
\begin{equation}\label{eq:euler}
  \vz_{\tau-\Delta\tau}
  = \vz_\tau + \Delta\tau \cdot f_\theta(\vu_{\mathrm{in}}, \vz_\tau, k, \tau),
  \quad \Delta\tau = 1/N.
\end{equation}
The resulting correction is deterministic conditioned on the perturbed state, avoiding the repeated stochastic denoising updates of DDPM-based refinement.
\Cref{alg:train} summarizes training; the inference procedure is given in \cref{alg:infer} in the appendix.

\FloatBarrier
\begin{algorithm}[!t]
\caption{FlowRefiner Training}\label{alg:train}
\begin{algorithmic}[1]\small
\REQUIRE Pairs $\{(\vu_{\mathrm{in}}, \vu_{\mathrm{gt}})\}$, steps $K$, schedule $\{\sigma_k\}$, prior $\pi$
\FOR{each batch $(\vu_{\mathrm{in}}, \vu_{\mathrm{gt}})$}
  \STATE Sample $k \sim \mathrm{Uniform}\{0,\ldots,K\}$
  \IF{$k = 0$}
    \STATE $\mathcal{L} \leftarrow \|f_\theta(\vu_{\mathrm{in}}, k{=}0) - \vu_{\mathrm{gt}}\|^2$
  \ELSE
    \STATE Sample $\tau \sim \mathrm{Uniform}[0,1]$, $\veps \sim \pi$
    \STATE $\vz_\tau \leftarrow (1-\tau)\,\vu_{\mathrm{gt}} + \tau\,\sigma_k\,\veps$
    \STATE $\mathcal{L} \leftarrow \|f_\theta(\vu_{\mathrm{in}}, \vz_\tau, k, \tau) - (-\veps)\|^2$
  \ENDIF
  \STATE Update $\theta$ via $\nabla_\theta \mathcal{L}$
\ENDFOR
\end{algorithmic}
\end{algorithm}
\FloatBarrier

\subsection{Decoupled Sigma Schedule.}\label{sec:schedule}

The role of the noise schedule in iterative refinement is to determine how correction is distributed across refinement steps.
Existing coupled schedules~\cite{karras2022elucidating} tie refinement depth to perturbation magnitude, so increasing the number of refinement steps can also enlarge the effective noise range.
For 3D turbulence, however, increasing the refinement depth $K$ should make correction more gradual, not make the refinement problem itself harder.
This motivates our decoupled schedule, which fixes the noise interval independently of $K$ and lets refinement depth control only how finely that interval is traversed.

We set $\sigma_{\max}{=}0.01$ and $\sigma_{\min}{=}0.001$, and distribute $K$ values log-uniformly:
\begin{equation}\label{eq:decoupled}
  \sigma_k
  = \sigma_{\max}
  \left(\frac{\sigma_{\min}}{\sigma_{\max}}\right)^{(k-1)/(K-1)}.
\end{equation}
Under this design, adding refinement steps subdivides the same small-noise regime more finely rather than expanding it.
Refinement depth is therefore decoupled from perturbation magnitude, directly addressing the noise--depth coupling described in \cref{sec:limitations}.

\begin{table*}[!t]
\centering
\caption{Autoregressive prediction on FIT isotropic turbulence ($128^3$). Per-channel RMSE ($\downarrow$) and SSIM ($\uparrow$) across three rollout rounds. Deterministic baselines from~\cite{dai2025pest}. \textbf{Bold}: best, \underline{underlined}: second best.}\label{tab:baselines}
\vspace{-0.2cm}
\resizebox{\textwidth}{!}{%
\setlength{\tabcolsep}{2.2pt}
\fontsize{7pt}{9pt}\selectfont
\begin{tabular}{l cccccc cccccc cccccc cccccc}
\toprule
& \multicolumn{6}{c}{$u$} & \multicolumn{6}{c}{\cellcolor{bandgray}$v$} & \multicolumn{6}{c}{$w$} & \multicolumn{6}{c}{\cellcolor{bandgray}$p$} \\
\cmidrule(lr){2-7} \cmidrule(lr){8-13} \cmidrule(lr){14-19} \cmidrule(lr){20-25}
& \multicolumn{3}{c}{RMSE $\downarrow$} & \multicolumn{3}{c}{SSIM $\uparrow$}
& \multicolumn{3}{c}{\cellcolor{bandgray}RMSE $\downarrow$} & \multicolumn{3}{c}{\cellcolor{bandgray}SSIM $\uparrow$}
& \multicolumn{3}{c}{RMSE $\downarrow$} & \multicolumn{3}{c}{SSIM $\uparrow$}
& \multicolumn{3}{c}{\cellcolor{bandgray}RMSE $\downarrow$} & \multicolumn{3}{c}{\cellcolor{bandgray}SSIM $\uparrow$} \\
Method & R1 & R2 & R3 & R1 & R2 & R3
  & \cellcolor{bandgray}R1 & \cellcolor{bandgray}R2 & \cellcolor{bandgray}R3 & \cellcolor{bandgray}R1 & \cellcolor{bandgray}R2 & \cellcolor{bandgray}R3
  & R1 & R2 & R3 & R1 & R2 & R3
  & \cellcolor{bandgray}R1 & \cellcolor{bandgray}R2 & \cellcolor{bandgray}R3 & \cellcolor{bandgray}R1 & \cellcolor{bandgray}R2 & \cellcolor{bandgray}R3 \\
\midrule
\multicolumn{25}{l}{\emph{Spectral operators}} \\
FNO3D & .163&.234&.265 & .625&.484&.415 & \cellcolor{bandgray}.161&\cellcolor{bandgray}.234&\cellcolor{bandgray}.265 & \cellcolor{bandgray}.690&\cellcolor{bandgray}.521&\cellcolor{bandgray}.445 & .162&.236&.269 & .633&.465&.394 & \cellcolor{bandgray}.086&\cellcolor{bandgray}.121&\cellcolor{bandgray}.141 & \cellcolor{bandgray}.599&\cellcolor{bandgray}.512&\cellcolor{bandgray}.474 \\
U-FNO & .153&.219&.259 & .659&.525&.437 & \cellcolor{bandgray}.150&\cellcolor{bandgray}.216&\cellcolor{bandgray}.255 & \cellcolor{bandgray}.717&\cellcolor{bandgray}.572&\cellcolor{bandgray}.482 & .153&.219&.256 & .645&.511&.429 & \cellcolor{bandgray}.083&\cellcolor{bandgray}.118&\cellcolor{bandgray}.139 & \cellcolor{bandgray}.632&\cellcolor{bandgray}.532&\cellcolor{bandgray}.482 \\
TFNO & .177&.283&.344 & .628&.425&.314 & \cellcolor{bandgray}.174&\cellcolor{bandgray}.284&\cellcolor{bandgray}.349 & \cellcolor{bandgray}.686&\cellcolor{bandgray}.446&\cellcolor{bandgray}.326 & .180&.286&.351 & .610&.398&.289 & \cellcolor{bandgray}.091&\cellcolor{bandgray}.157&\cellcolor{bandgray}.205 & \cellcolor{bandgray}.582&\cellcolor{bandgray}.472&\cellcolor{bandgray}.395 \\
\midrule
\multicolumn{25}{l}{\emph{Attention-based}} \\
Transolver & .211&.231&.257 & .469&.536&.424 & \cellcolor{bandgray}.213&\cellcolor{bandgray}.232&\cellcolor{bandgray}.258 & \cellcolor{bandgray}.459&\cellcolor{bandgray}.521&\cellcolor{bandgray}.424 & .212&.232&.258 & .447&.504&.404 & \cellcolor{bandgray}.117&\cellcolor{bandgray}.128&\cellcolor{bandgray}.147 & \cellcolor{bandgray}.617&\cellcolor{bandgray}.684&\cellcolor{bandgray}.591 \\
DPOT & .203&.226&.253 & .658&.577&.498 & \cellcolor{bandgray}.205&\cellcolor{bandgray}.228&\cellcolor{bandgray}.257 & \cellcolor{bandgray}.629&\cellcolor{bandgray}.557&\cellcolor{bandgray}.487 & .204&.228&.256 & .618&.541&.468 & \cellcolor{bandgray}.112&\cellcolor{bandgray}.125&\cellcolor{bandgray}.146 & \cellcolor{bandgray}.764&\cellcolor{bandgray}.705&\cellcolor{bandgray}.637 \\
FactFormer & .171&.213&.252 & .755&.626&.508 & \cellcolor{bandgray}.171&\cellcolor{bandgray}.213&\cellcolor{bandgray}.252 & \cellcolor{bandgray}.734&\cellcolor{bandgray}.611&\cellcolor{bandgray}.503 & .171&.213&.251 & .729&.600&.487 & \cellcolor{bandgray}.092&\cellcolor{bandgray}.114&\cellcolor{bandgray}.139 & \cellcolor{bandgray}.820&\cellcolor{bandgray}.734&\cellcolor{bandgray}.649 \\
\midrule
\multicolumn{25}{l}{\emph{Physics-informed}} \\
PINO & .163&.234&.265 & .630&.485&.415 & \cellcolor{bandgray}.162&\cellcolor{bandgray}.235&\cellcolor{bandgray}.267 & \cellcolor{bandgray}.678&\cellcolor{bandgray}.509&\cellcolor{bandgray}.432 & .163&.236&.268 & .628&.469&.400 & \cellcolor{bandgray}.088&\cellcolor{bandgray}.121&\cellcolor{bandgray}.139 & \cellcolor{bandgray}.574&\cellcolor{bandgray}.503&\cellcolor{bandgray}.472 \\
PEST & \underline{.127}&.165&.191 & .878&.796&.745 & \cellcolor{bandgray}\underline{.128}&\cellcolor{bandgray}.174&\cellcolor{bandgray}.208 & \cellcolor{bandgray}.880&\cellcolor{bandgray}.779&\cellcolor{bandgray}.715 & \underline{.128}&.173&.206 & .873&.774&.708 & \cellcolor{bandgray}\underline{.060}&\cellcolor{bandgray}.086&\cellcolor{bandgray}.104 & \cellcolor{bandgray}.922&\cellcolor{bandgray}.905&\cellcolor{bandgray}.881 \\
\midrule
\multicolumn{25}{l}{\emph{Generative}} \\
Cond.\ FM & .128&.163&.182 & \underline{.919}&\underline{.864}&\underline{.829} & \cellcolor{bandgray}.135&\cellcolor{bandgray}.170&\cellcolor{bandgray}\underline{.188} & \cellcolor{bandgray}\underline{.916}&\cellcolor{bandgray}\underline{.864}&\cellcolor{bandgray}\underline{.831} & .131&.166&\underline{.185} & \underline{.916}&\underline{.860}&\underline{.822} & \cellcolor{bandgray}.063&\cellcolor{bandgray}\underline{.084}&\cellcolor{bandgray}\underline{.098} & \cellcolor{bandgray}\underline{.962}&\cellcolor{bandgray}\underline{.934}&\cellcolor{bandgray}\underline{.907} \\
PDE-Ref. & .131&\underline{.159}&\underline{.180} & .911&\underline{.864}&.825 & \cellcolor{bandgray}.138&\cellcolor{bandgray}\underline{.168}&\cellcolor{bandgray}.189 & \cellcolor{bandgray}.908&\cellcolor{bandgray}.862&\cellcolor{bandgray}.823 & .135&\underline{.164}&\underline{.185} & .908&\underline{.860}&.819 & \cellcolor{bandgray}.067&\cellcolor{bandgray}.091&\cellcolor{bandgray}.113 & \cellcolor{bandgray}.959&\cellcolor{bandgray}.923&\cellcolor{bandgray}.876 \\
DiT-Ref. & .177&.212&.244 & .831&.752&.669 & \cellcolor{bandgray}.185&\cellcolor{bandgray}.220&\cellcolor{bandgray}.253 & \cellcolor{bandgray}.828&\cellcolor{bandgray}.755&\cellcolor{bandgray}.677 & .180&.215&.248 & .828&.748&.665 & \cellcolor{bandgray}.089&\cellcolor{bandgray}.110&\cellcolor{bandgray}.132 & \cellcolor{bandgray}.925&\cellcolor{bandgray}.892&\cellcolor{bandgray}.843 \\
\textbf{FlowRefiner} & \textbf{.077}&\textbf{.101}&\textbf{.117} & \textbf{.971}&\textbf{.950}&\textbf{.933} & \cellcolor{bandgray}\textbf{.078}&\cellcolor{bandgray}\textbf{.103}&\cellcolor{bandgray}\textbf{.118} & \cellcolor{bandgray}\textbf{.972}&\cellcolor{bandgray}\textbf{.952}&\cellcolor{bandgray}\textbf{.936} & \textbf{.077}&\textbf{.102}&\textbf{.118} & \textbf{.971}&\textbf{.950}&\textbf{.933} & \cellcolor{bandgray}\textbf{.035}&\cellcolor{bandgray}\textbf{.051}&\cellcolor{bandgray}\textbf{.065} & \cellcolor{bandgray}\textbf{.989}&\cellcolor{bandgray}\textbf{.976}&\cellcolor{bandgray}\textbf{.958} \\
\bottomrule
\end{tabular}}%
\end{table*}

\subsection{Noise Design Space.}\label{sec:noise}

Because FM does not constrain the noise prior to be Gaussian, it opens a design axis unavailable under DDPM-based refinement.
We therefore study whether turbulence-aware perturbation priors can further improve refinement quality.

Most priors are generated in Fourier space by applying a frequency-dependent weight $w(\vk)$ to white Gaussian noise and then normalizing to unit variance.
We evaluate eight priors spanning distinct spectral profiles:
white noise as the default baseline;
blue noise with $w(k)\propto k^\beta$ at $\beta\in\{0.5, 1.0\}$, which suppresses low frequencies;
spectrum-matched and inverse-spectrum priors aligned with or inverted from the turbulence energy spectrum;
error-weighted noise shaped by the base model's per-frequency prediction error;
the von K\'{a}rm\'{a}n prior derived from the classical turbulence energy spectrum;
and divergence-free noise obtained by Leray projection of white noise.

\subsection{Physical Projection under FM.}\label{sec:projection}

Incompressible flow requires $\nabla \cdot \vu = 0$.
The Leray projection enforces this constraint by removing the irrotational component in Fourier space:
\begin{equation}\label{eq:leray}
  \hat{\vu}_{\mathrm{proj}}(\vk)
  = \hat{\vu}(\vk) - \frac{\vk\,(\vk \cdot \hat{\vu}(\vk))}{|\vk|^2},
\end{equation}
at $\mathcal{O}(N \log N)$ cost via one FFT pair.

A key difference between diffusion and FM is that DDPM's Gaussian forward process commutes with this projection, whereas FM's linear interpolation generally leaves the divergence-free subspace when the noise prior is unconstrained~\cite{christopher2024pgdm,pcdm2025,projectgenerate2025,utkarsh2025physicsconstrained}.
We therefore examine five projection strategies:
\emph{Vanilla} (no projection),
\emph{Terminal} (project only the final output),
\emph{PCFM}~\cite{utkarsh2025physicsconstrained} (per-step shooting and projection),
\emph{CCFM} (time-adaptive constraint tightening),
and \emph{Hard} (project every ODE substep); see \cref{tab:projection}.
This analysis probes the trade-off between physical admissibility and predictive accuracy under FM-based refinement.

\section{Experiments.}\label{sec:experiments}
\begin{figure*}[!tbp]
\centering
\includegraphics[width=\textwidth]{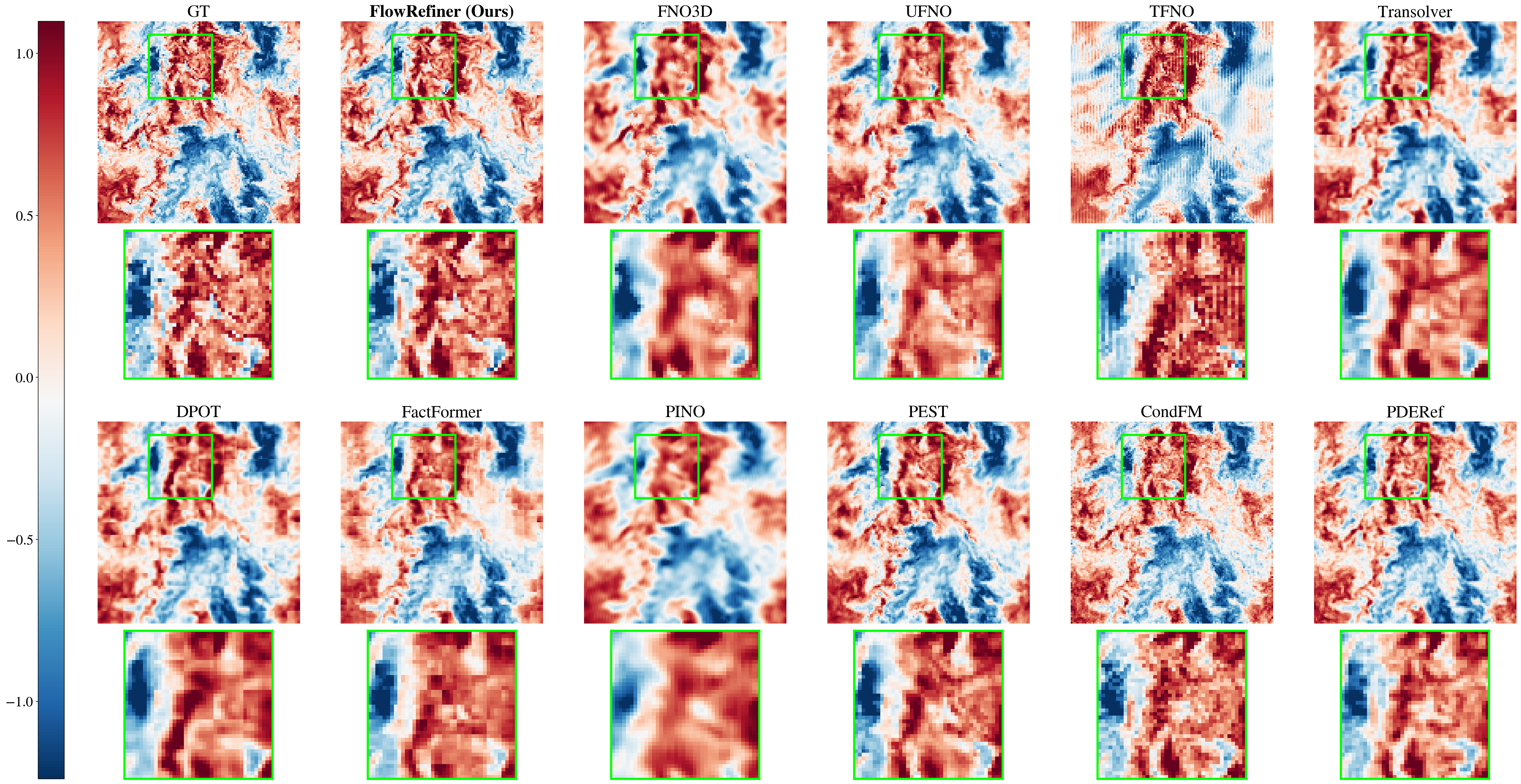}
\caption{$u$-component predictions on FIT at Round~3, showing the $z{=}60$ slice with magnified insets over a high-frequency region.}\label{fig:qual}
\end{figure*}

We organize the empirical study around four research questions:
\begin{itemize}[leftmargin=1.0em, itemsep=1pt, topsep=1pt, parsep=0pt, partopsep=0pt]
  \item \textbf{RQ1:} Does FlowRefiner outperform deterministic, physics-informed, and generative baselines?
  \item \textbf{RQ2:} Does the proposed refinement remain effective over long autoregressive rollouts?
  \item \textbf{RQ3:} Is the proposed decoupled small-noise regime essential, and how do refinement depth and ODE substeps affect performance?
  \item \textbf{RQ4:} What do projection strategies, perturbation priors, and reduced temporal context reveal about the robustness and design space of FlowRefiner?
\end{itemize}

\begin{figure*}[!t]
\centering
\includegraphics[width=\textwidth]{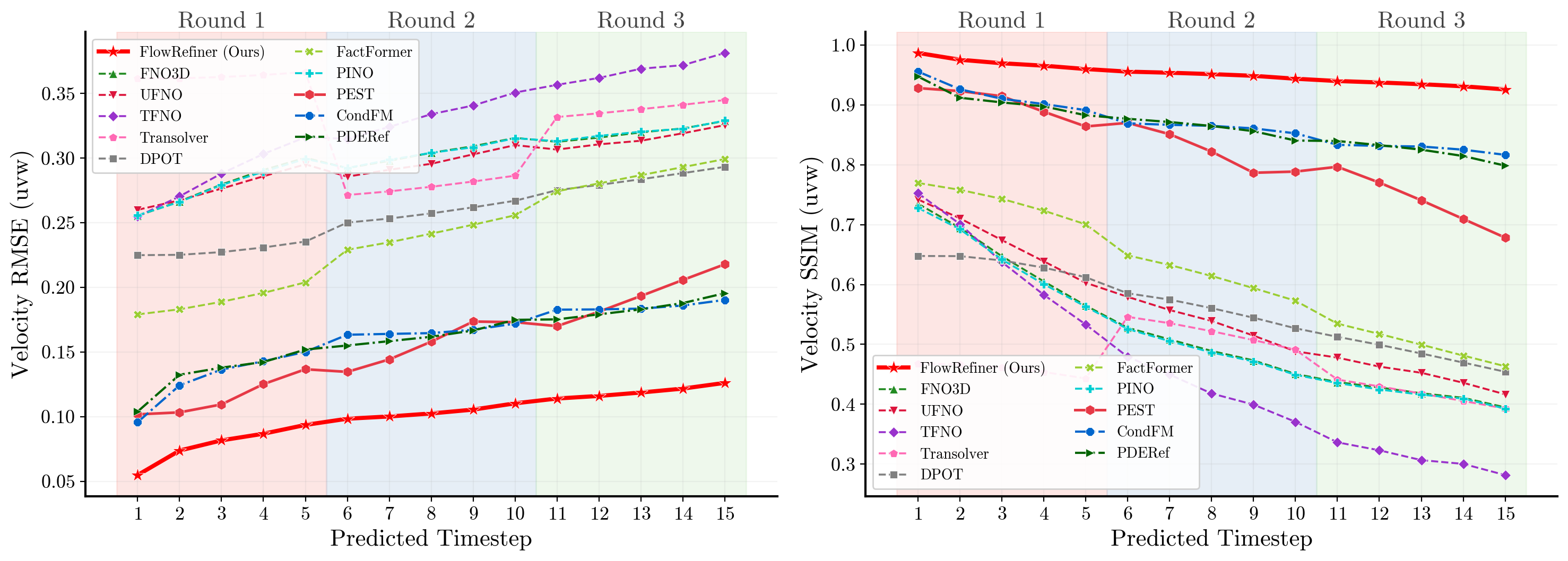}
\vspace{-0.7cm}
\caption{Per-timestep velocity RMSE ($\downarrow$, left) and SSIM ($\uparrow$, right) on FIT across 15 autoregressive steps (3 rounds). FlowRefiner (red) maintains the lowest RMSE and highest SSIM throughout the entire rollout horizon.}\label{fig:pertimestep}
\end{figure*}

\begin{figure*}[!tbp]
\centering
\includegraphics[width=\textwidth]{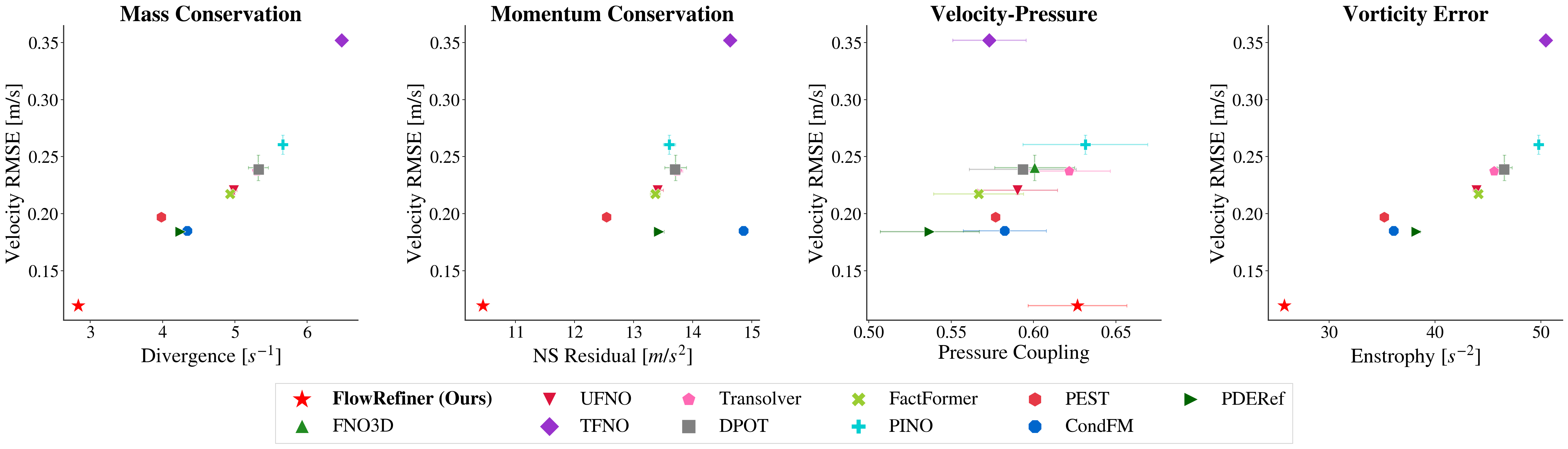}
\vspace{-0.6cm}
\caption{Physical consistency on FIT at Round~3. Velocity RMSE vs.\ four physics metrics; lower-left is optimal.}\label{fig:physics}
\end{figure*}

\subsection{Experimental Setup.}\label{sec:setup} We here introduce the common experimental setup, covering the datasets, model, and training configuration, evaluation metrics, and autoregressive rollout setting.

\noindent \textbf{Datasets.}
We evaluate on two 3D turbulence benchmarks with deliberately contrasting spectral characteristics.
\textbf{FIT}~\cite{li2008jhtdb,perlman2007jhtdb}, our primary benchmark, is forced isotropic turbulence from JHTDB at resolution $128^3$ with 4 channels ($u,v,w,p$)---fully developed turbulence with abundant high-frequency details and fine-scale structures, the regime where spectral bias is most severe.
\textbf{TGV}, the Taylor--Green vortex at $64{\times}128{\times}128$ with the same 4 channels, serves as a robustness check in the opposite regime: it starts smooth ($w{=}0$ at $t{=}0$) and remains largely low-frequency within our evaluation horizon (see \cref{app:spectra} for the quantitative energy-spectrum comparison with FIT and \cref{tab:tgv_baselines} for full per-channel numbers), testing whether FlowRefiner stays competitive outside the high-frequency regime it targets.
Unless otherwise noted, we use five input frames to predict the next five frames per rollout round; the effect of reducing temporal context is ablated in \cref{sec:exp_noise}. All fields are normalized to zero mean and unit variance per channel before training; metrics are computed after denormalization to physical scales.

\noindent \textbf{Architecture and training.}
We use RefinerUNet3D with hidden dimension 64, channel multipliers $(1,2,2,4)$, and 2 residual blocks per level, totaling 50.4M parameters.
Training uses AdamW~\cite{loshchilov2019adamw} with learning rate $10^{-4}$, weight decay $10^{-5}$, and cosine annealing to $10^{-6}$.
We apply exponential moving average (EMA) with decay 0.995 and gradient checkpointing for memory efficiency.
Inference uses $N{=}2$ Euler ODE substeps per refinement step.

\noindent \textbf{Metrics.}
We report the root mean squared error (RMSE) and the structural similarity index (SSIM)~\cite{wang2004image}, with the SSIM calculated using a $7\times7\times7$ 3D window. For autoregressive evaluation on FIT, we report per-round RMSE across 3 rollout rounds.

\subsection{Main Results.}\label{sec:exp_baselines}

\Cref{tab:baselines} shows that on FIT, FlowRefiner achieves the lowest per-channel RMSE across all rollout rounds, outperforming PEST by 40\% at R1. This advantage is consistent across all four channels and is further supported by the qualitative, per-timestep, and physics-consistency results below, which indicate stronger long-horizon prediction in fully developed 3D turbulence.
As a supplementary robustness check, we also evaluate on TGV, a largely laminar, low-frequency regime that lies outside the main setting targeted by FlowRefiner. There, FlowRefiner remains best on the dominant $u$ and $v$ components, while PEST and FactFormer perform better on the weaker $w$ and $p$ channels. Full per-channel numbers are reported in \cref{tab:tgv_baselines} (\cref{app:spectra}).


\noindent \textbf{Per-timestep results.}
\Cref{fig:qual} visualizes $u$-component predictions at autoregressive Round~3, after 15 prediction steps, the regime where baseline degradation is most pronounced.
FlowRefiner produces the closest match to the DNS ground truth, preserving sharp velocity interfaces and fine-grained turbulent filaments in the magnified inset.
Conditional FM, the closest competitor, captures large-scale patterns but exhibits noise speckles and blurred boundaries from generating directly from noise.
Deterministic neural operators (FNO3D, U-FNO, TFNO, PINO) show strong over-smoothing by R3; attention-based methods (Transolver, DPOT, FactFormer) retain more large-scale structure but still lose fine features.
PEST, the strongest deterministic baseline, shows mild over-smoothing compared with FlowRefiner.
Additional qualitative comparisons ($v$, $w$, $p$) are in \cref{app:qual_extra}.

\noindent \textbf{Per-timestep performance.}
\Cref{fig:pertimestep} tracks velocity RMSE and SSIM across all 15 autoregressive steps on FIT.
FlowRefiner maintains a consistent gap over all baselines: its SSIM stays above 0.93 while the best baseline (PEST) drops from 0.88 to 0.75 by the final step.
The curves reveal two degradation patterns: deterministic baselines show smooth, steady error growth, whereas generative baselines (Cond.\ FM, PDE-Refiner) exhibit step-wise jumps at round boundaries.
FlowRefiner's smoother degradation indicates that predict-then-refine with small-sigma FM avoids both failure modes.

\noindent \textbf{Physical consistency without explicit supervision.}
\Cref{fig:physics} evaluates predictions against four physics metrics at Round~3 on FIT: divergence, NS residual, pressure coupling, and vorticity error.
FlowRefiner achieves the lowest velocity RMSE across all four panels while maintaining physical residuals comparable to PEST, which uses explicit Navier--Stokes and divergence constraints.
This suggests that FM refinement on a strong base predictor implicitly learns physically consistent dynamics.

\begin{table}[!tbp]
\centering
\caption{ODE substeps $N$ ablation. Differences beyond $N{=}2$ are within $10^{-4}$ RMSE (numerical-precision regime); we use $N{=}2$ throughout. \textbf{Bold}: best per column.}\label{tab:ode_n}
\footnotesize
\begin{tabular}{@{}l cccc@{}}
\toprule
& \multicolumn{2}{c}{FIT ($K{=}2$)} & \multicolumn{2}{c}{TGV ($K{=}4$)} \\
\cmidrule(lr){2-3} \cmidrule(lr){4-5}
$N$ & R1 & R3 & R1 & R2 \\
\midrule
1  & .0709 & .1085 & .0589 & .0793 \\
2  & .0693 & .1060 & .0571 & .0776 \\
3  & .0693 & .1060 & .0571 & .0776 \\
5  & \textbf{.0692} & .1060 & .0570 & \textbf{.0774} \\
10 & \textbf{.0692} & \textbf{.1059} & \textbf{.0569} & .0775 \\
\bottomrule
\end{tabular}
\end{table}

\begin{table}[!tbp]
\centering
\caption{Sigma range ablation on FIT ($K{=}2$, $N{=}1$). Larger $\sigma_{\max}$ consistently degrades accuracy.}\label{tab:sigma_range}
\vspace{-0.2cm}
\footnotesize
\begin{tabular}{@{}lccc@{}}
\toprule
$\sigma_{\max}$ (\% signal) & R1 & R2 & R3 \\
\midrule
0.01 (${\sim}$1\%) & \textbf{.0709} & \textbf{.0933} & \textbf{.1085} \\
0.03 (${\sim}$3\%) & .0749 & .0995 & .1155 \\
0.05 (${\sim}$5\%) & .0810 & .1085 & .1247 \\
\bottomrule
\end{tabular}
\end{table}

\subsection{Refinement Steps and Sigma Schedule.}\label{sec:exp_main}

We examine how refinement performance depends on the ODE discretization, sigma range, refinement depth $K$, and schedule design. \Cref{tab:ode_n} shows that under the decoupled schedule, $N{=}2$ ODE substeps already suffice on both datasets: $N{=}1$ incurs a small penalty (2.3\% R1 on FIT at $K{=}2$; 3.2\% R1 on TGV at $K{=}4$), while $N{\geq}3$ brings negligible further gain. \Cref{tab:sigma_range} further shows that refinement works best in a strict small-noise regime. On FIT, increasing $\sigma_{\max}$ from 0.01 to 0.05 degrades R1 by 14\%, and on TGV, larger values can cause catastrophic failure. These results support using a small fixed perturbation range with $N{=}2$ in all main experiments.

We then study the effect of refinement depth $K$ on FIT under the proposed decoupled schedule. As shown in \Cref{tab:jhu_main}, iterative refinement is consistently helpful when additional steps subdivide a fixed perturbation interval rather than enlarge it. Compared with the base predictor ($K{=}0$), using $K{=}2$ reduces R1 RMSE from 0.0772 to 0.0709, with similar gains at R2 and R3. $K{=}4$ yields nearly identical accuracy (R1 0.0712), indicating that depth selection is robust within this range; performance then plateaus with mild fluctuations at larger $K$ but remains better than the base predictor at all depths tested, suggesting that once the correction path is sufficiently resolved, extra refinement steps add computation without yielding further benefit.

\Cref{tab:jhu_main} also compares the proposed decoupled schedule with alternative schedules, each reported at its best $K$. Linear, cprod, and cprod-large all perform worse than both the decoupled schedule and the unrefined base model. This shows that the main issue is not refinement itself, but schedule design. When the perturbation magnitude grows with refinement depth, early refinement steps leave the small-noise regime and become destructive on $128^3$ turbulence. Quantitatively, the decoupled schedule at $K{=}2$ outperforms the best coupled configuration (cprod, $K{=}1$, R1 0.0900) by 21.2\% on R1, compared with an 8.2\% gain over the base predictor. This confirms that decoupling the perturbation range from the refinement depth is essential for stable refinement in 3D turbulence. Using the best setting together with $N{=}2$ ODE substeps improves R1 to 0.0693, corresponding to a 10.2\% gain over the base predictor.

\begin{table}[!tbp]
\centering
\caption{$K$ ablation on FIT ($128^3$) with $N{=}1$. AR RMSE across 3 rounds. For non-decoupled schedules, we report each at its best $K$.}\label{tab:jhu_main}
\vspace{-0.1cm}
\footnotesize
\begin{tabular}{@{}lccc@{}}
\toprule
Configuration & R1 & R2 & R3 \\
\midrule
$K{=}0$ & .0772 & .1030 & .1207 \\
\midrule
\multicolumn{4}{@{}l}{\emph{Decoupled}} \\
$K{=}2$ & \textbf{.0709} & \textbf{.0933} & \textbf{.1085} \\
$K{=}4$ & .0712 & .0939 & .1095 \\
$K{=}6$ & .0744 & .0978 & .1140 \\
$K{=}8$ & .0736 & .0963 & .1118 \\
\midrule
\multicolumn{4}{@{}l}{\emph{Other schedules (each at its best $K$)}} \\
Linear ($K{=}1$)        & .0921 & .1211 & .1411 \\
Cprod ($K{=}1$)         & .0900 & .1142 & .1335 \\
Cprod-large ($K{=}1$)   & .0919 & .1215 & .1381 \\
\bottomrule
\end{tabular}
\end{table}

\subsection{Projection Strategies.}\label{sec:exp_projection}

We evaluate five projection strategies (\cref{sec:projection}) on the same checkpoint without retraining.
All constrained strategies reduce divergence to ${\sim}10^{-5}$ (near machine precision), but at severe accuracy cost: 3$\times$ on TGV ($u$-RMSE 0.022$\to$0.067) and 4.3$\times$ on FIT (0.084$\to$0.360).
All five strategies yield nearly identical degradation, suggesting the cost is intrinsic to unconstrained FM paths rather than an artifact of any particular enforcement method.
Constrained methods do exhibit slower AR growth (${\sim}$+30\%/round vs.\ +141\%), but the initial accuracy gap is not recovered.
This is consistent with~\cite{projectgenerate2025}: output projection alone is insufficient when the noise prior lies outside the constraint manifold.
Full per-strategy numbers are provided in \cref{app:projection}.

\subsection{Noise Priors and Robustness Analyses.}\label{sec:exp_noise}

\noindent \textbf{Noise priors.}
We compare the eight noise priors described in \cref{sec:noise} (full results in \cref{app:noise}). The overall variation is only $\pm$1--7\% RMSE, substantially smaller than the effect of $K$ or sigma schedule design, indicating that noise type is a minor tuning knob rather than a primary design decision.

\noindent \textbf{Input timesteps.}
\Cref{tab:input_ts} compares 5-frame and 1-frame input on FIT: reducing to a single frame increases R1 RMSE by 8\% but the gap narrows to 2.3\% by R3, and single-frame FlowRefiner still outperforms all baselines in \cref{tab:baselines}. That said, the strong temporal correlation of turbulent flows means a single frame already carries much of the short-horizon signal, so this likely understates the value of temporal context in regimes with weaker correlation.

\noindent \textbf{Autoregressive stability.}
As shown in \cref{tab:baselines,tab:jhu_main}, FlowRefiner maintains its advantage through 3 rollout rounds with error growth of $\sim$52\% (R1$\to$R3), comparable to the base predictor's 56\%. Refinement primarily improves the quality of initial predictions rather than fundamentally altering the dynamics of error propagation.

\begin{table}[!tbp]
\centering
\caption{Input timesteps ablation on FIT ($K{=}2$). Notation $a{\to}b$ denotes $a$ input frames predicting $b$ output frames per rollout round.}\label{tab:input_ts}
\vspace{-0.1cm}
\footnotesize
\begin{tabular}{@{}lccc@{}}
\toprule
Input & R1 & R2 & R3 \\
\midrule
$5{\to}5$ (default) & \textbf{.0709} & \textbf{.0933} & \textbf{.1085} \\
$1{\to}5$ & .0766 & .0963 & .1110 \\
\bottomrule
\end{tabular}
\end{table}

\section{Discussion and Conclusion.}\label{sec:discussion}

In this paper, we presented FlowRefiner, an FM-based iterative refinement framework for 3D turbulent flow simulation. We identify a diffusion--refinement mismatch in DDPM-style post-hoc correction for 3D turbulence: refinement depth is coupled to perturbation magnitude, the shared network is trained with discontinuous targets across stages, and stochastic perturbations accumulate along an otherwise deterministic autoregressive rollout. FlowRefiner resolves these limitations in a unified way. Its decoupled sigma schedule fixes the perturbation range independently of refinement depth, so increasing $K$ makes correction more gradual rather than more destructive; its FM-based formulation replaces the stage-wise switch from solution regression to noise regression with a single correction objective shared across all refinement stages; and its deterministic ODE-based update avoids the repeated stochastic denoising steps that otherwise inject forward-accumulating variance. Overall, these design choices make iterative correction more stable and better matched to complex 3D turbulent flows. Empirically, this leads to state-of-the-art autoregressive prediction with strong physical consistency, while our ablations further show that perturbation-prior design is secondary to the refinement mechanism itself and that physical projection under FM introduces a fundamental accuracy--consistency trade-off. In conclusion, these findings suggest that FM is a more suitable foundation for post-hoc refinement in deterministic multiscale dynamics. We hope this work motivates broader task-specific and application-driven instantiations of this refinement paradigm across scientific forecasting and other multiscale modeling domains.

\bibliographystyle{siamplain}
\bibliography{references}

@article{lippe2024pderefiner,
  title={Pde-refiner: Achieving accurate long rollouts with neural pde solvers},
  author={Lippe, Phillip and Veeling, Bas and Perdikaris, Paris and Turner, Richard and Brandstetter, Johannes},
  journal={Advances in Neural Information Processing Systems},
  volume={36},
  pages={67398--67433},
  year={2023}
}

@inproceedings{lipman2023flow,
  title={Flow Matching for Generative Modeling},
  author={Lipman, Yaron and Chen, Ricky TQ and Ben-Hamu, Heli and Nickel, Maximilian and Le, Matthew},
  booktitle={The Eleventh International Conference on Learning Representations}
}

@inproceedings{liu2023rectified,
  title={Flow Straight and Fast: Learning to Generate and Transfer Data with Rectified Flow},
  author={Liu, Xingchao and Gong, Chengyue and others},
  booktitle={The Eleventh International Conference on Learning Representations}
}

@article{projectgenerate2025,
  title={Project and Generate: Divergence-Free Neural Operators for Incompressible Flows},
  author={Li, Xigui and Zhang, Hongwei and Jiang, Ruoxi and Chen, Deshu and Lin, Chensen and Han, Limei and Qi, Yuan and Guo, Xin and Cheng, Yuan},
  journal={arXiv preprint arXiv:2603.24500},
  year={2026}
}

@article{wu2025latentfm,
  title={Generative latent neural pde solver using flow matching},
  author={Li, Zijie and Zhou, Anthony and Farimani, Amir Barati},
  journal={arXiv preprint arXiv:2503.22600},
  year={2025}
}

@article{fourierflow2025,
  title={FourierFlow: Frequency-aware Flow Matching for Generative Turbulence Modeling},
  author={Wang, Haixin and Pan, Jiashu and Wu, Hao and Zhang, Fan and Wu, Tailin},
  journal={arXiv preprint arXiv:2506.00862},
  year={2025}
}

@article{residualfm2024,
  title={Flow matching Operators for Residual-Augmented Probabilistic Learning of Partial Differential Equations},
  author={Bhola, Sahil and Duraisamy, Karthik},
  journal={arXiv preprint arXiv:2512.12749},
  year={2025}
}

@article{cfo2025,
  title={CFO: Learning Continuous-Time PDE Dynamics via Flow-Matched Neural Operators},
  author={Hou, Xianglong and Huang, Xinquan and Perdikaris, Paris},
  journal={arXiv preprint arXiv:2512.05297},
  year={2025}
}

@inproceedings{pdespectralrefiner2025,
  title={PDESpectralRefiner: Achieving More Accurate Long Rollouts with Spectral Adjustment},
  author={Luo, Li},
  booktitle={2025 4th International Symposium on Computer Applications and Information Technology (ISCAIT)},
  pages={2097--2100},
  year={2025},
  organization={IEEE}
}

@article{kohl2024acdm,
  title={Benchmarking autoregressive conditional diffusion models for turbulent flow simulation},
  author={Kohl, Georg and Chen, Li-Wei and Thuerey, Nils},
  journal={Neural Networks},
  pages={108641},
  year={2026},
  publisher={Elsevier}
}

@article{pcdm2025,
  title={Physics-Constrained Diffusion Model for Synthesis of 3D Turbulent Data},
  author={Li, Tianyi and Buzzicotti, Michele and Bonaccorso, Fabio and Biferale, Luca},
  journal={arXiv preprint arXiv:2603.12834},
  year={2026}
}

@article{utkarsh2025physicsconstrained,
  title={Physics-constrained flow matching: Sampling generative models with hard constraints},
  author={Utkarsh, Utkarsh and Cai, Pengfei and Edelman, Alan and Gomez-Bombarelli, Rafael and Rackauckas, Christopher Vincent},
  journal={arXiv preprint arXiv:2506.04171},
  year={2025}
}

@article{christopher2024pgdm,
  title={Constrained synthesis with projected diffusion models},
  author={Christopher, Jacob K and Baek, Stephen and Fioretto, Ferdinando},
  journal={Advances in Neural Information Processing Systems},
  volume={37},
  pages={89307--89333},
  year={2024}
}

@article{li2021fno,
  title={Fourier neural operator for parametric partial differential equations},
  author={Li, Zongyi and Kovachki, Nikola and Azizzadenesheli, Kamyar and Liu, Burigede and Bhattacharya, Kaushik and Stuart, Andrew and Anandkumar, Anima},
  journal={arXiv preprint arXiv:2010.08895},
  year={2020}
}

@inproceedings{tran2023ffno,
  title={Factorized Fourier Neural Operators},
  author={Tran, Alasdair and Mathews, Alexander and Xie, Lexing and Ong, Cheng Soon},
  booktitle={The Eleventh International Conference on Learning Representations}
}

@article{gupta2022modernunet,
  title={Towards Multi-spatiotemporal-scale Generalized PDE Modeling},
  author={Gupta, Jayesh K and Brandstetter, Johannes},
  journal={Transactions on Machine Learning Research}
}

@article{
li2023oformer,
title={Transformer for Partial Differential Equations{\textquoteright} Operator Learning},
author={Zijie Li and Kazem Meidani and Amir Barati Farimani},
journal={Transactions on Machine Learning Research},
issn={2835-8856},
year={2023},
url={https://openreview.net/forum?id=EPPqt3uERT},
note={}
}

@article{li2008jhtdb,
  title={A public turbulence database cluster and applications to study Lagrangian evolution of velocity increments in turbulence},
  author={Li, Yi and Perlman, Eric and Wan, Minping and Yang, Yunke and Meneveau, Charles and Burns, Randal and Chen, Shiyi and Szalay, Alexander and Eyink, Gregory},
  journal={Journal of Turbulence},
  number={9},
  pages={N31},
  year={2008},
  publisher={Taylor \& Francis}
}

@inproceedings{perlman2007jhtdb,
  title={Data exploration of turbulence simulations using a database cluster},
  author={Perlman, Eric and Burns, Randal and Li, Yi and Meneveau, Charles},
  booktitle={Proceedings of the 2007 ACM/IEEE Conference on Supercomputing},
  pages={1--11},
  year={2007}
}

@article{ho2020ddpm,
  title={Denoising diffusion probabilistic models},
  author={Ho, Jonathan and Jain, Ajay and Abbeel, Pieter},
  journal={Advances in neural information processing systems},
  volume={33},
  pages={6840--6851},
  year={2020}
}

@article{kingma2024diffusionmeetsflow,
  title={Understanding diffusion objectives as the elbo with simple data augmentation},
  author={Kingma, Diederik and Gao, Ruiqi},
  journal={Advances in Neural Information Processing Systems},
  volume={36},
  pages={65484--65516},
  year={2023}
}

@inproceedings{
loshchilov2019adamw,
title={Decoupled Weight Decay Regularization},
author={Ilya Loshchilov and Frank Hutter},
booktitle={International Conference on Learning Representations},
year={2019},
url={https://openreview.net/forum?id=Bkg6RiCqY7},
}

@article{albergo2023stochastic,
  title={Stochastic interpolants: A unifying framework for flows and diffusions},
  author={Albergo, Michael and Boffi, Nicholas M and Vanden-Eijnden, Eric},
  journal={Journal of Machine Learning Research},
  volume={26},
  number={209},
  pages={1--80},
  year={2025}
}

@article{raissi2019pinn,
  title={Physics-informed neural networks: A deep learning framework for solving forward and inverse problems involving nonlinear partial differential equations},
  author={Raissi, Maziar and Perdikaris, Paris and Karniadakis, George E},
  journal={Journal of Computational physics},
  volume={378},
  pages={686--707},
  year={2019},
  publisher={Elsevier}
}

@article{wang2021gradient,
  title={Understanding and mitigating gradient flow pathologies in physics-informed neural networks},
  author={Wang, Sifan and Teng, Yujun and Perdikaris, Paris},
  journal={SIAM Journal on Scientific Computing},
  volume={43},
  number={5},
  pages={A3055--A3081},
  year={2021},
  publisher={SIAM}
}

@article{wang2024causal,
  title={Respecting causality for training physics-informed neural networks},
  author={Wang, Sifan and Sankaran, Shyam and Perdikaris, Paris},
  journal={Computer Methods in Applied Mechanics and Engineering},
  volume={421},
  pages={116813},
  year={2024},
  publisher={Elsevier}
}

@inproceedings{
wu2024transolver,
title={Transolver: A Fast Transformer Solver for {PDE}s on General Geometries},
author={Haixu Wu and Huakun Luo and Haowen Wang and Jianmin Wang and Mingsheng Long},
booktitle={Forty-first International Conference on Machine Learning},
year={2024},
url={https://openreview.net/forum?id=Ywl6pODXjB}
}

@article{dai2025pest,
  title={{PEST}: Physics-Enhanced Swin Transformer for 3{D} Turbulence Simulation},
  author={Dai, Yilong and Chen, Shengyu and Jia, Xiaowei and Givi, Peyman and Yu, Runlong},
  journal={arXiv preprint arXiv:2602.10150},
  year={2026}
}

@article{du2024confild,
  title={Conditional neural field latent diffusion model for generating spatiotemporal turbulence},
  author={Du, Pan and Parikh, Meet Hemant and Fan, Xiantao and Liu, Xin-Yang and Wang, Jian-Xun},
  journal={Nature Communications},
  volume={15},
  number={1},
  pages={10416},
  year={2024},
  publisher={Nature Publishing Group UK London}
}

@misc{dai2026flowlearnerspdesphysicstophysics,
      title={Flow Learners for PDEs: Toward a Physics-to-Physics Paradigm for Scientific Computing}, 
      author={Yilong Dai and Shengyu Chen and Xiaowei Jia and Runlong Yu},
      year={2026},
      eprint={2604.07366},
      archivePrefix={arXiv},
      primaryClass={cs.LG},
      url={https://arxiv.org/abs/2604.07366}, 
}

@misc{dai2026learningpdesolversphysics,
      title={Learning PDE Solvers with Physics and Data: A Unifying View of Physics-Informed Neural Networks and Neural Operators}, 
      author={Yilong Dai and Shengyu Chen and Ziyi Wang and Xiaowei Jia and Yiqun Xie and Vipin Kumar and Runlong Yu},
      year={2026},
      eprint={2601.14517},
      archivePrefix={arXiv},
      primaryClass={cs.LG},
}

@article{salimans2022progressive,
  title={Progressive distillation for fast sampling of diffusion models},
  author={Salimans, Tim and Ho, Jonathan},
  journal={arXiv preprint arXiv:2202.00512},
  year={2022}
}

@inproceedings{
lienen2024from,
title={From Zero to Turbulence: Generative Modeling for 3D Flow Simulation},
author={Marten Lienen and David L{\"u}dke and Jan Hansen-Palmus and Stephan G{\"u}nnemann},
booktitle={The Twelfth International Conference on Learning Representations},
year={2024},
url={https://openreview.net/forum?id=ZhlwoC1XaN}
}

@inproceedings{rahaman2019spectral,
  title={On the spectral bias of neural networks},
  author={Rahaman, Nasim and Baratin, Aristide and Arpit, Devansh and Draxler, Felix and Lin, Min and Hamprecht, Fred and Bengio, Yoshua and Courville, Aaron},
  booktitle={International conference on machine learning},
  pages={5301--5310},
  year={2019},
  organization={PMLR}
}

@article{takamoto2022pdebench,
  title={Pdebench: An extensive benchmark for scientific machine learning},
  author={Takamoto, Makoto and Praditia, Timothy and Leiteritz, Raphael and MacKinlay, Daniel and Alesiani, Francesco and Pfl{\"u}ger, Dirk and Niepert, Mathias},
  journal={Advances in neural information processing systems},
  volume={35},
  pages={1596--1611},
  year={2022}
}

@article{wen2022u,
  title={U-FNO—An enhanced Fourier neural operator-based deep-learning model for multiphase flow},
  author={Wen, Gege and Li, Zongyi and Azizzadenesheli, Kamyar and Anandkumar, Anima and Benson, Sally M},
  journal={Advances in Water Resources},
  volume={163},
  pages={104180},
  year={2022},
  publisher={Elsevier}
}

@inproceedings{
hao2024dpot,
title={{DPOT}: Auto-Regressive Denoising Operator Transformer for Large-Scale {PDE} Pre-Training},
author={Zhongkai Hao and Chang Su and Songming Liu and Julius Berner and Chengyang Ying and Hang Su and Anima Anandkumar and Jian Song and Jun Zhu},
booktitle={Forty-first International Conference on Machine Learning},
year={2024},
url={https://openreview.net/forum?id=X7UnDevHOM}
}

@article{li2023scalable,
  title={Scalable transformer for pde surrogate modeling},
  author={Li, Zijie and Shu, Dule and Barati Farimani, Amir},
  journal={Advances in Neural Information Processing Systems},
  volume={36},
  pages={28010--28039},
  year={2023}
}

@article{li2024physics,
  title={Physics-informed neural operator for learning partial differential equations},
  author={Li, Zongyi and Zheng, Hongkai and Kovachki, Nikola and Jin, David and Chen, Haoxuan and Liu, Burigede and Azizzadenesheli, Kamyar and Anandkumar, Anima},
  journal={ACM/IMS Journal of Data Science},
  volume={1},
  number={3},
  pages={1--27},
  year={2024},
  publisher={ACM New York, NY}
}

@article{wang2004image,
  title={Image quality assessment: from error visibility to structural similarity},
  author={Wang, Zhou and Bovik, Alan C and Sheikh, Hamid R and Simoncelli, Eero P},
  journal={IEEE transactions on image processing},
  volume={13},
  number={4},
  pages={600--612},
  year={2004},
  publisher={IEEE}
}

@inproceedings{
brandstetter2022message,
title={Message Passing Neural {PDE} Solvers},
author={Johannes Brandstetter and Daniel E. Worrall and Max Welling},
booktitle={International Conference on Learning Representations},
year={2022},
url={https://openreview.net/forum?id=vSix3HPYKSU}
}

@article{oommen2024integrating,
  title={Integrating neural operators with diffusion models improves spectral representation in turbulence modeling},
  author={Oommen, Vivek and Bora, Aniruddha and Zhang, Zhen and Karniadakis, George Em},
  journal={arXiv preprint arXiv:2409.08477},
  year={2024}
}

@inproceedings{
song2021scorebased,
title={Score-Based Generative Modeling through Stochastic Differential Equations},
author={Yang Song and Jascha Sohl-Dickstein and Diederik P Kingma and Abhishek Kumar and Stefano Ermon and Ben Poole},
booktitle={International Conference on Learning Representations},
year={2021},
url={https://openreview.net/forum?id=PxTIG12RRHS}
}

@article{shysheya2024conditional,
  title={On conditional diffusion models for pde simulations},
  author={Shysheya, Aliaksandra and Diaconu, Cristiana and Bergamin, Federico and Perdikaris, Paris and Hern{\'a}ndez-Lobato, Jos{\'e} M and Turner, Richard E and Mathieu, Emile},
  journal={Advances in Neural Information Processing Systems},
  volume={37},
  pages={23246--23300},
  year={2024}
}

@article{
yoo2026diffusionrollout,
title={DiffusionRollout: Uncertainty-Aware Rollout Planning in Long-Horizon {PDE} Solving},
author={Seungwoo Yoo and Juil Koo and Daehyeon Choi and Minhyuk Sung},
journal={Transactions on Machine Learning Research},
issn={2835-8856},
year={2026},
url={https://openreview.net/forum?id=OCzcGOzgzz},
note={}
}

@article{karras2022elucidating,
  title={Elucidating the design space of diffusion-based generative models},
  author={Karras, Tero and Aittala, Miika and Aila, Timo and Laine, Samuli},
  journal={Advances in neural information processing systems},
  volume={35},
  pages={26565--26577},
  year={2022}
}

\clearpage

\appendix
\section{FlowRefiner Inference Procedure.}\label{app:inference}

\Cref{alg:infer} details the inference procedure of FlowRefiner: starting from the deterministic base prediction, $K$ refinement steps are applied, each integrating the learned FM velocity field through $N$ Euler substeps. The inference procedure mirrors the training loop (\cref{alg:train}) but differs in two scheduling aspects: (i) the refinement index $k$ is traversed sequentially from $1$ to $K$ at inference, whereas training uniformly samples a single $k \in \{0,\ldots,K\}$ per batch; (ii) the FM interpolation parameter $\tau$ is uniformly sampled during training, while inference integrates the ODE from $\tau{=}1$ to $\tau{=}0$ through $N$ equally spaced Euler substeps within each refinement step. The same network $f_\theta$ is shared across all roles.

\begin{algorithm}[t]
\caption{FlowRefiner Inference}\label{alg:infer}
\begin{algorithmic}[1]
\REQUIRE Input frames $\vu_{\mathrm{in}}$, model $f_\theta$, schedule $\{\sigma_k\}_{k=1}^{K}$, ODE substeps $N$, noise prior $\pi$
\ENSURE Refined prediction $\hat{\vu}$
\STATE $\hat{\vu} \leftarrow f_\theta(\vu_{\mathrm{in}}, k{=}0)$ \hfill$\triangleright$ Base prediction
\FOR{$k = 1$ to $K$}
  \STATE Sample $\veps \sim \pi$
  \STATE $\vz \leftarrow \hat{\vu} + \sigma_k \veps$ \hfill$\triangleright$ Inject noise
  \FOR{$n = 0$ to $N{-}1$}
    \STATE $\tau \leftarrow 1 - n/N$
    \STATE $\vz \leftarrow \vz + \tfrac{1}{N}\, f_\theta(\vu_{\mathrm{in}}, \vz, k, \tau)$
  \ENDFOR
  \STATE $\hat{\vu} \leftarrow \vz$
\ENDFOR
\RETURN $\hat{\vu}$
\end{algorithmic}
\end{algorithm}

\noindent\textbf{Compute cost per block.}
For a single autoregressive block of $T_{\mathrm{out}}$ output frames, \cref{alg:infer} executes one base-prediction forward pass followed by $K$ refinement steps with $N$ Euler substeps each, for a total of $1 + K \cdot N$ network forwards. Under the main configuration ($K{=}2$, $N{=}2$) this amounts to $5$ UNet forwards per block---roughly $5\times$ the base predictor's inference cost. The deeper $K{=}4$, $N{=}2$ configuration costs $9$ forwards per block. These counts are consistent with the ablations in \cref{sec:exp_main}: \cref{tab:ode_n} shows that $N{=}2$ already attains the full benefit of ODE integration and $N{\geq}3$ yields negligible further gains, and \cref{tab:jhu_main} shows that $K{=}2$ is sufficient under the decoupled schedule. The overhead of FM-based refinement at inference is therefore fixed and modest, and is not a function of either the sigma schedule shape or the noise prior.

\section{Physical Projection: Per-Strategy Numbers.}\label{app:projection}

\Cref{tab:projection} reports the full per-strategy results for the physical projection analysis of \cref{sec:exp_projection}.
All constrained strategies achieve near-machine-precision divergence ($\text{div}_{\max} {\sim} 10^{-5}$) but incur a 3--4$\times$ accuracy degradation that is essentially identical across methods.

\begin{table}[t]
\centering
\caption{Physical projection results on a fixed FlowRefiner checkpoint (no retraining). Vanilla = no projection; all other strategies reduce divergence to machine precision but degrade accuracy. AR Gr.: relative RMSE growth per rollout round.}\label{tab:projection}
\footnotesize
\begin{tabular}{@{}lcccc@{}}
\toprule
& \multicolumn{2}{c}{RMSE} & & \\
\cmidrule(lr){2-3}
Strategy & TGV ($u$) & FIT & $\text{div}_{\max}$ & AR Gr. \\
\midrule
Vanilla & 0.022 & 0.084 & $\sim$0.1 & +141\% \\
Terminal proj. & 0.067 & 0.360 & $\sim 10^{-5}$ & +30\% \\
PCFM & 0.067 & 0.361 & $\sim 10^{-5}$ & +31\% \\
CCFM & 0.067 & 0.360 & $\sim 10^{-5}$ & +30\% \\
Hard proj. & 0.068 & 0.362 & $\sim 10^{-5}$ & +29\% \\
\bottomrule
\end{tabular}
\end{table}

\noindent\textbf{All constrained strategies collapse to the same point.} Terminal projection, PCFM, CCFM, and Hard projection produce TGV $u$-RMSE in the range $0.067$--$0.068$ and FIT RMSE in $0.360$--$0.362$; the differences between them are under $1\%$. The four strategies differ substantially in where and how often they enforce $\nabla \cdot \vu = 0$ along the ODE path (only at the end, at each integrator shoot, with time-adaptive tightening, or at every substep), yet the resulting accuracy is indistinguishable. This supports the structural reading of \cref{sec:exp_projection}: the cost is not a tuning artifact of any particular enforcement scheme but an intrinsic property of FM paths. Linear interpolation $\vz_\tau = (1-\tau)\vu_{\mathrm{gt}} + \tau\sigma_k\veps$ generally leaves the divergence-free subspace when $\veps$ is not itself divergence-free, so any projection back onto that subspace must pay the same initial-state correction cost regardless of its timing.

\noindent\textbf{Accuracy--admissibility trade-off.} The two regimes in \cref{tab:projection} sit on opposite ends of a trade-off curve. Vanilla starts from a small FIT RMSE of $0.084$ but grows at $+141\%$ per rollout round; every constrained variant starts from a $4.3\times$ larger RMSE of $\sim 0.36$ but grows at only $\sim{+}30\%$ per round. Crucially, these rates are compounded on very different bases, so comparing ``$+141\%$ vs.\ $+30\%$'' directly is misleading: within our 3-round horizon the two curves do not cross, and projection slows error growth without ever recovering the initial $4.3\times$ accuracy gap. The implication is that physical admissibility and predictive accuracy are not separable objectives under unconstrained FM refinement, as already noted in \cref{sec:exp_projection}. This is also consistent with the noise-prior evidence in \cref{app:noise}: the div-free white prior does not meaningfully beat white ($+0.5\%$ on FIT), so shaping the prior to live on the constraint manifold is not, on its own, sufficient either---the constraint must be built into the refinement formulation, not bolted on.

\section{Dataset Spectral Characteristics.}\label{app:spectra}

\begin{figure}[t]
\centering
\includegraphics[width=\columnwidth]{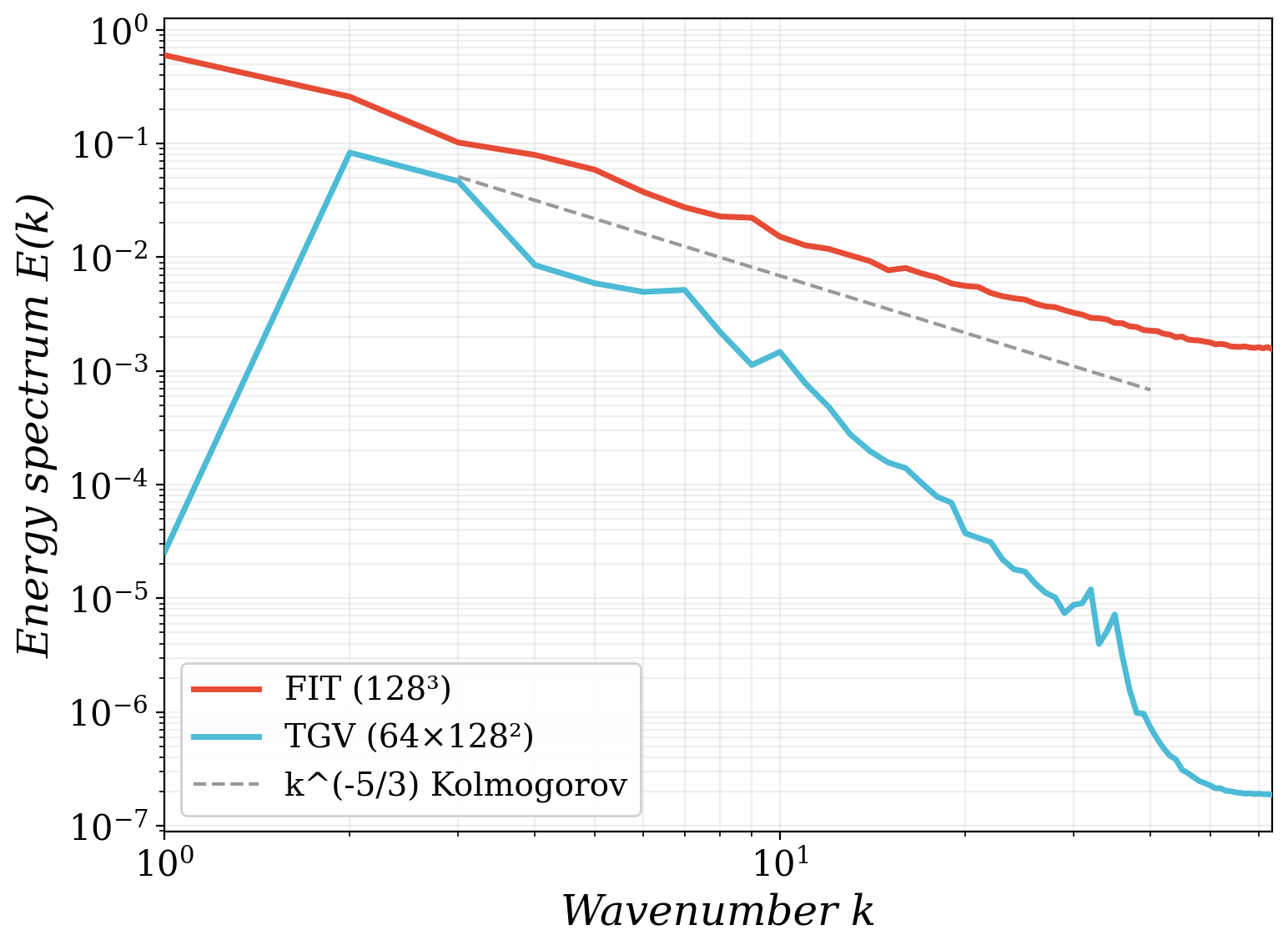}
\caption{Angle-averaged velocity energy spectra $E(k)$ for FIT ($128^3$) and TGV ($64{\times}128^2$). FIT follows the Kolmogorov $k^{-5/3}$ slope across a broad inertial range, while TGV concentrates energy in low-wavenumber modes and decays sharply at higher $k$.}
\label{fig:spectra}
\end{figure}

\Cref{fig:spectra} compares the angle-averaged velocity energy spectra $E(k)$ of FIT and TGV, supporting the dataset positioning in \cref{sec:setup}. FIT decays roughly along the Kolmogorov $k^{-5/3}$ slope across more than a decade of wavenumbers, retaining substantial energy out to $k \sim 50$---the signature of fully developed turbulence with a broad inertial cascade. TGV, in contrast, concentrates its energy in a narrow peak near $k{=}2$ and drops by nearly five orders of magnitude by $k{=}40$, consistent with a largely laminar flow dominated by a few large-scale coherent modes. This order-of-magnitude spectral separation is why the two benchmarks probe complementary aspects of the method: FIT stresses high-frequency recovery (the regime FlowRefiner targets), while TGV tests whether the method remains competitive when high-frequency content is nearly absent.

Quantitatively, the two spectra diverge progressively with $k$: the ratio $E_{\mathrm{FIT}}(k)/E_{\mathrm{TGV}}(k)$ grows from order unity near $k{=}3$ to roughly $10^{2}$ by $k{\approx}10$ and to $10^{3}$ or more by $k{\approx}30$--$50$. FIT's inertial-range segment spans nearly a full decade in $k$ along the Kolmogorov slope, while TGV exhibits no well-defined Kolmogorov-style inertial range and instead drops nearly monotonically once past its narrow low-$k$ peak. Almost all of TGV's energy within our evaluation horizon therefore lives in a handful of coherent modes at $k \in [1,3]$. This separation motivates our treatment of the two benchmarks asymmetrically in \cref{sec:exp_baselines}: FIT is the primary target, and TGV is reported as a robustness check in a regime where the mechanism FlowRefiner is designed to exploit (small-noise correction of under-resolved high-frequency structure) has limited room to act.

Because TGV lies outside the high-frequency regime FlowRefiner is designed for, we include it only as a supplementary robustness check and report its full per-channel numbers in \cref{tab:tgv_baselines} rather than in the main text. Consistent with the narrative in \cref{sec:exp_baselines}, FlowRefiner is clearly best on the dominant $u$ and $v$ velocity components, while PEST and FactFormer lead on the weaker $w$ and $p$ channels, where explicit Navier--Stokes supervision provides a stronger inductive bias in this near-laminar regime.

\noindent\textbf{Per-channel breakdown.} On the energy-carrying components $u$ and $v$, FlowRefiner nearly halves the R1 RMSE relative to the strongest baseline PEST (0.014 vs.\ 0.026, a 46\% relative reduction) and lifts SSIM from 0.951 to 0.989. At R2 the gap narrows but remains in FlowRefiner's favor (0.033 vs.\ 0.036; SSIM 0.940 vs.\ 0.909). On the weak components, however, the ordering reverses: PEST achieves 0.024 on $w$ versus FlowRefiner's 0.089 ($\sim$3.7$\times$), and 0.010 on $p$ versus 0.021 ($\sim$2$\times$). This asymmetry is structural rather than incidental. TGV is initialized with $w \equiv 0$ and remains dominated by a single planar shear pattern over our evaluation horizon, so $w$ and $p$ stay in a low-energy, smooth regime where hard physics constraints (divergence-free velocity, Poisson-consistent pressure) provide an informative inductive bias. PEST encodes exactly these constraints, whereas FlowRefiner's advantage is concentrated on correcting under-resolved high-frequency structure, which is largely absent in TGV's $w$ and $p$ fields. The TGV result therefore does not contradict the main FIT finding; it delineates when FM-based small-noise refinement is and is not the dominant source of error reduction.

\begin{table*}[!tbp]
\centering
\caption{Autoregressive prediction on TGV ($64{\times}128{\times}128$). Per-channel RMSE ($\downarrow$) and SSIM ($\uparrow$) across two rounds. Baselines from~\cite{dai2025pest}. \textbf{Bold}: best, \underline{underlined}: second best.}\label{tab:tgv_baselines}
\resizebox{0.75\textwidth}{!}{%
\setlength{\tabcolsep}{2.2pt}
\fontsize{7pt}{9pt}\selectfont
\begin{tabular}{l cccc cccc cccc cccc}
\toprule
& \multicolumn{4}{c}{$u$} & \multicolumn{4}{c}{\cellcolor{bandgray}$v$} & \multicolumn{4}{c}{$w$} & \multicolumn{4}{c}{\cellcolor{bandgray}$p$} \\
\cmidrule(lr){2-5} \cmidrule(lr){6-9} \cmidrule(lr){10-13} \cmidrule(lr){14-17}
& \multicolumn{2}{c}{RMSE $\downarrow$} & \multicolumn{2}{c}{SSIM $\uparrow$}
& \multicolumn{2}{c}{\cellcolor{bandgray}RMSE $\downarrow$} & \multicolumn{2}{c}{\cellcolor{bandgray}SSIM $\uparrow$}
& \multicolumn{2}{c}{RMSE $\downarrow$} & \multicolumn{2}{c}{SSIM $\uparrow$}
& \multicolumn{2}{c}{\cellcolor{bandgray}RMSE $\downarrow$} & \multicolumn{2}{c}{\cellcolor{bandgray}SSIM $\uparrow$} \\
Method & R1 & R2 & R1 & R2
  & \cellcolor{bandgray}R1 & \cellcolor{bandgray}R2 & \cellcolor{bandgray}R1 & \cellcolor{bandgray}R2
  & R1 & R2 & R1 & R2
  & \cellcolor{bandgray}R1 & \cellcolor{bandgray}R2 & \cellcolor{bandgray}R1 & \cellcolor{bandgray}R2 \\
\midrule
\multicolumn{17}{l}{\emph{Spectral operators}} \\
FNO3D & .040&.066 & .830&.685 & \cellcolor{bandgray}.038&\cellcolor{bandgray}.066 & \cellcolor{bandgray}.853&\cellcolor{bandgray}.695 & .065&.073 & .531&.454 & \cellcolor{bandgray}.062&\cellcolor{bandgray}.062 & \cellcolor{bandgray}.461&\cellcolor{bandgray}.385 \\
U-FNO & .038&.061 & .850&.708 & \cellcolor{bandgray}.038&\cellcolor{bandgray}.062 & \cellcolor{bandgray}.851&\cellcolor{bandgray}.696 & .065&.076 & .541&.420 & \cellcolor{bandgray}.054&\cellcolor{bandgray}.056 & \cellcolor{bandgray}.534&\cellcolor{bandgray}.424 \\
TFNO & .036&.086 & .872&.632 & \cellcolor{bandgray}.037&\cellcolor{bandgray}.085 & \cellcolor{bandgray}.866&\cellcolor{bandgray}.637 & .050&.077 & .666&.414 & \cellcolor{bandgray}.036&\cellcolor{bandgray}.048 & \cellcolor{bandgray}.661&\cellcolor{bandgray}.462 \\
\midrule
\multicolumn{17}{l}{\emph{Attention-based}} \\
Transolver & .036&.051 & .883&.790 & \cellcolor{bandgray}.036&\cellcolor{bandgray}.051 & \cellcolor{bandgray}.884&\cellcolor{bandgray}.791 & .045&.058 & \underline{.767}&.664 & \cellcolor{bandgray}.021&\cellcolor{bandgray}.026 & \cellcolor{bandgray}.809&\cellcolor{bandgray}.700 \\
DPOT & .038&.055 & .867&.766 & \cellcolor{bandgray}.038&\cellcolor{bandgray}.055 & \cellcolor{bandgray}.867&\cellcolor{bandgray}.768 & .046&.056 & .762&.658 & \cellcolor{bandgray}.022&\cellcolor{bandgray}.027 & \cellcolor{bandgray}.790&\cellcolor{bandgray}.660 \\
FactFormer & .031&.049 & .892&.787 & \cellcolor{bandgray}.031&\cellcolor{bandgray}.049 & \cellcolor{bandgray}.893&\cellcolor{bandgray}.787 & \underline{.040}&\underline{.055} & \textbf{.826}&\underline{.687} & \cellcolor{bandgray}\underline{.020}&\cellcolor{bandgray}\underline{.025} & \cellcolor{bandgray}.788&\cellcolor{bandgray}.670 \\
\midrule
\multicolumn{17}{l}{\emph{Physics-informed}} \\
PINO & .041&.064 & .831&.694 & \cellcolor{bandgray}.041&\cellcolor{bandgray}.066 & \cellcolor{bandgray}.821&\cellcolor{bandgray}.681 & .068&.077 & .543&.421 & \cellcolor{bandgray}.060&\cellcolor{bandgray}.060 & \cellcolor{bandgray}.473&\cellcolor{bandgray}.399 \\
PEST & \underline{.026}&\underline{.036} & \underline{.951}&\underline{.909} & \cellcolor{bandgray}\underline{.026}&\cellcolor{bandgray}\underline{.036} & \cellcolor{bandgray}\underline{.951}&\cellcolor{bandgray}\underline{.908} & \textbf{.024}&\textbf{.026} & .764&\textbf{.743} & \cellcolor{bandgray}\textbf{.010}&\cellcolor{bandgray}\textbf{.013} & \cellcolor{bandgray}\textbf{.970}&\cellcolor{bandgray}\textbf{.942} \\
\midrule
\multicolumn{17}{l}{\emph{Generative}} \\
Cond.\ FM & .045&.068 & .848&.723 & \cellcolor{bandgray}.045&\cellcolor{bandgray}.067 & \cellcolor{bandgray}.847&\cellcolor{bandgray}.729 & .091&.103 & .442&.325 & \cellcolor{bandgray}.026&\cellcolor{bandgray}.039 & \cellcolor{bandgray}.709&\cellcolor{bandgray}.526 \\
PDE-Ref. & .033&.058 & .914&.792 & \cellcolor{bandgray}.033&\cellcolor{bandgray}.058 & \cellcolor{bandgray}.916&\cellcolor{bandgray}.792 & .090&.104 & .384&.283 & \cellcolor{bandgray}.025&\cellcolor{bandgray}.036 & \cellcolor{bandgray}.773&\cellcolor{bandgray}.613 \\
DiT-Ref. & .039&.059 & .900&.813 & \cellcolor{bandgray}.039&\cellcolor{bandgray}.058 & \cellcolor{bandgray}.901&\cellcolor{bandgray}.816 & .092&.106 & .395&.283 & \cellcolor{bandgray}.027&\cellcolor{bandgray}.042 & \cellcolor{bandgray}.745&\cellcolor{bandgray}.549 \\
\textbf{FlowRefiner} & \textbf{.014}&\textbf{.033} & \textbf{.989}&\textbf{.940} & \cellcolor{bandgray}\textbf{.014}&\cellcolor{bandgray}\textbf{.033} & \cellcolor{bandgray}\textbf{.989}&\cellcolor{bandgray}\textbf{.941} & .089&.101 & .436&.352 & \cellcolor{bandgray}.021&\cellcolor{bandgray}.036 & \cellcolor{bandgray}\underline{.876}&\cellcolor{bandgray}\underline{.721} \\
\bottomrule
\end{tabular}}%
\end{table*}

\section{Additional Qualitative Comparisons.}\label{app:qual_extra}

\begin{figure*}[!tbp]
\centering
\includegraphics[width=\textwidth]{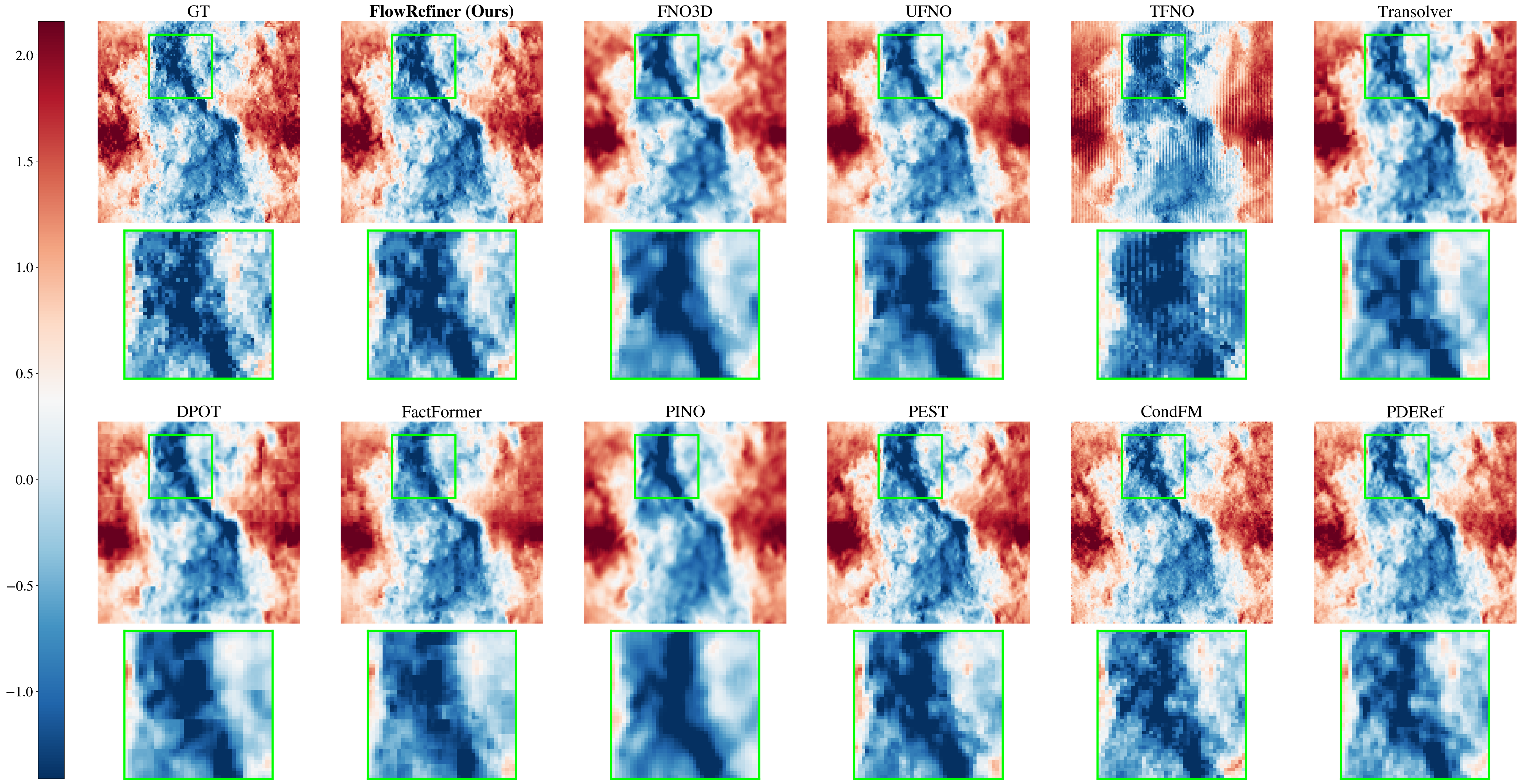}
\caption{$v$-component predictions on FIT at autoregressive Round~3 ($t{=}6.0$), with magnified insets over a high-frequency turbulent region. Layout as in \cref{fig:qual}.}\label{fig:qual_v}
\end{figure*}

\begin{figure*}[!tbp]
\centering
\includegraphics[width=\textwidth]{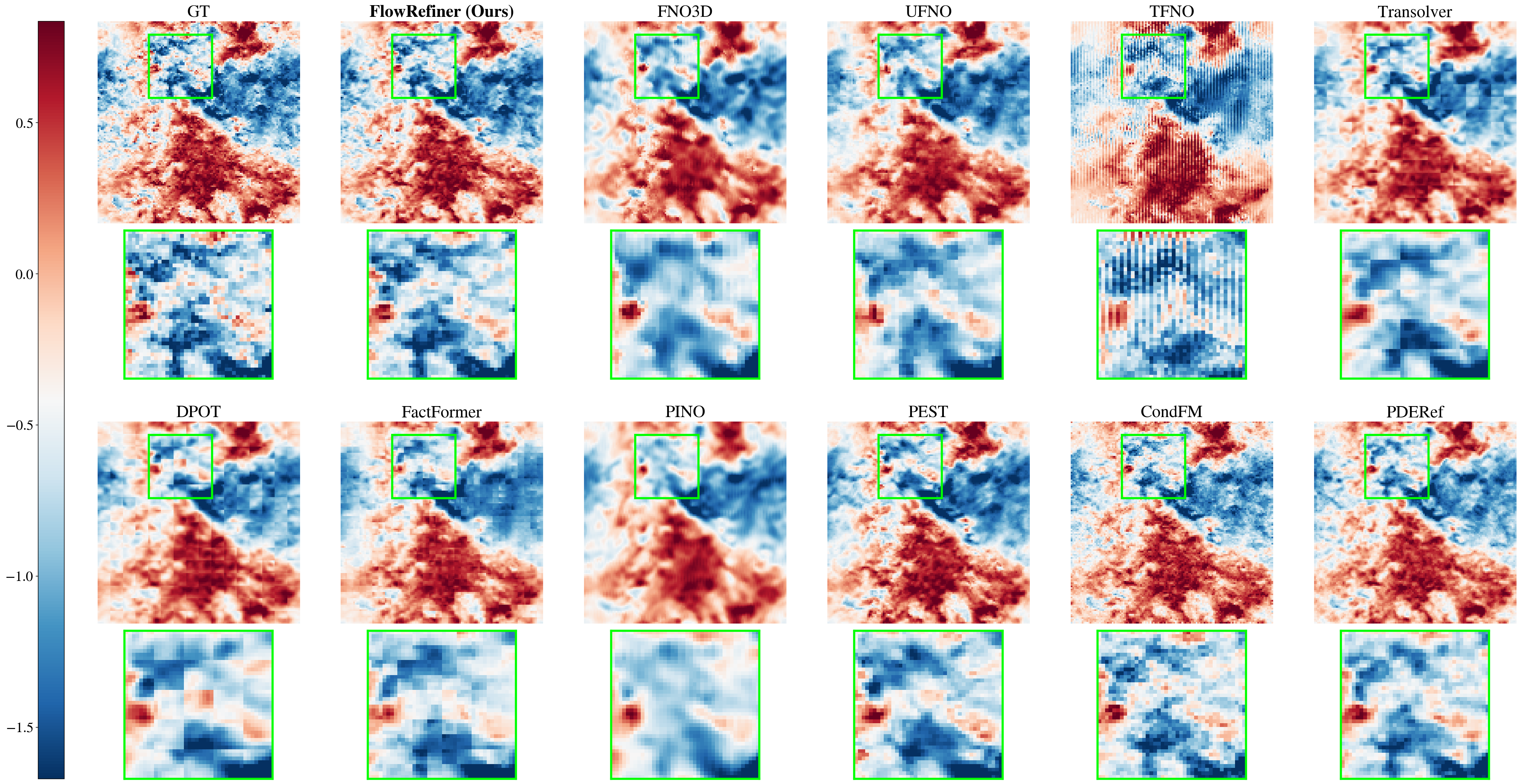}
\caption{$w$-component predictions on FIT at autoregressive Round~3. Layout as in \cref{fig:qual}.}\label{fig:qual_w}
\end{figure*}

\begin{figure*}[!tbp]
\centering
\includegraphics[width=\textwidth]{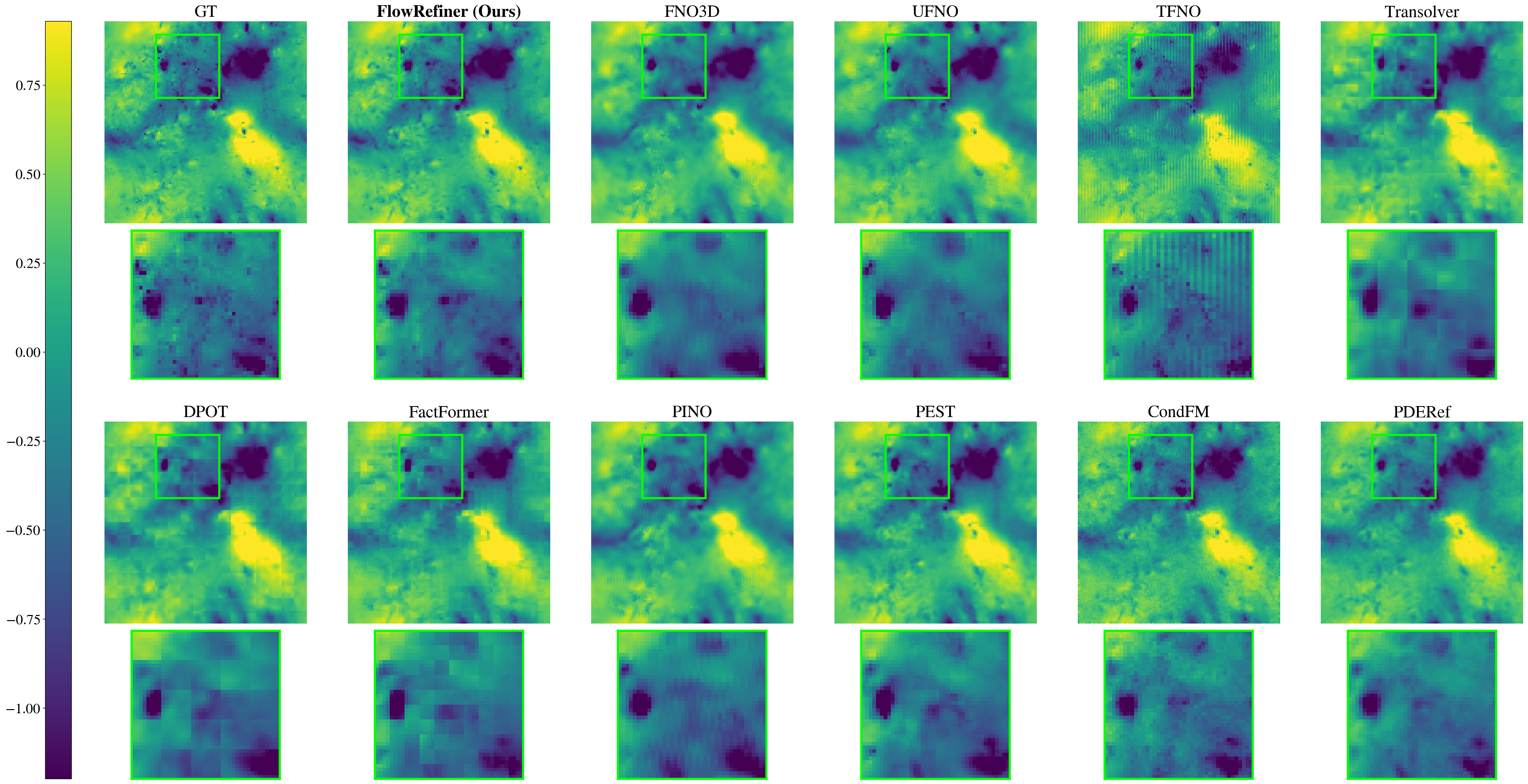}
\caption{Pressure ($p$) predictions on FIT at autoregressive Round~3. Layout as in \cref{fig:qual}.}\label{fig:qual_p}
\end{figure*}

\Cref{fig:qual_v,fig:qual_w,fig:qual_p} provide qualitative comparisons for the remaining channels ($v$, $w$, and pressure $p$) on FIT at autoregressive Round~3, complementing the $u$-channel results in \cref{fig:qual}.
Across all four channels, FlowRefiner consistently preserves sharp interfaces and fine-scale turbulent structures within the zoomed regions, whereas most baselines exhibit over-smoothing, structural degradation, or high-frequency noise.

\noindent\textbf{$v$-component (Fig.~\ref{fig:qual_v}).}
The $v$ field at R3 shows a predominantly blue (negative $v$) slice with interspersed red (positive $v$) patches and a clear slanted gradient. FlowRefiner's prediction preserves the sharpness of this gradient and the secondary filamentary structure inside the zoom window, matching GT most closely. The spectral operators FNO3D, U-FNO, and PINO exhibit strong over-smoothing in the zoom, with their fine-scale structure largely washed out. TFNO additionally shows a noticeable vertical-stripe pattern superimposed on the field in the zoom region. Attention-based models (Transolver, DPOT, FactFormer) retain more of the large-scale interface than the spectral operators but still lose the sharper sub-features visible in GT. Among the generative baselines, Cond.\ FM produces a visibly blurred boundary, and PDE-Ref introduces granular high-frequency noise that is not present in GT.

\noindent\textbf{$w$-component (Fig.~\ref{fig:qual_w}).}
The $w$ field contains a richer set of small-scale turbulent filaments than $u$ or $v$ at R3. FlowRefiner's prediction follows the filament geometry of GT most closely within the zoom. FNO3D, U-FNO, and PINO again show pronounced over-smoothing. TFNO's zoom region carries a strong vertical-stripe artifact that is inconsistent with the underlying turbulent structure. Transolver, DPOT, and FactFormer keep the large-scale layout but collapse the filaments into broader blobs. PDE-Ref introduces clearly visible speckle-like high-frequency noise in the zoom, and Cond.\ FM shows a milder, smeared-textural version of the same kind of noise---a failure mode distinct from the operators' over-smoothing but equally visible against GT.

\noindent\textbf{Pressure $p$ (Fig.~\ref{fig:qual_p}).}
The pressure field is dominated by a few large-scale high-pressure and low-pressure lobes, with notably weaker small-scale content than any of the velocity components. FlowRefiner reproduces the position, amplitude, and smooth transitions of these lobes faithfully, and its zoom region matches GT with no visible artifact. TFNO once again exhibits a vertical-stripe pattern inside the zoom that is clearly inconsistent with the smooth pressure field. The remaining baselines recover lobe positions correctly, but with varying degrees of over-smoothing in the transitions between them.

\noindent\textbf{Summary across channels.}
FlowRefiner is the only method across all four channels ($u$ in \cref{fig:qual}, $v$/$w$/$p$ here) that simultaneously (i) preserves the large-scale structure visible outside the zoom, (ii) recovers fine-scale details inside the zoom, and (iii) avoids both the over-smoothing of the deterministic operators and the high-frequency speckle of the diffusion/FM generative baselines. It also does not display the vertical-stripe artifact that TFNO exhibits in multiple channels. This visually confirms the dual-failure-mode observation in \cref{sec:exp_baselines} that predict-then-refine with small-sigma FM avoids the two typical error signatures of baseline approaches.

\section{Noise Type Ablation: Full Results.}\label{app:noise}

\Cref{tab:noise} reports the full noise type comparison across both datasets. Before discussing the results, we give in \cref{tab:noise_defs} a compact reference of the eight priors studied, each generated by shaping white Gaussian noise with a frequency-dependent weight $w(\vk)$ and then normalizing the sample to unit variance.

\begin{table}[t]
\centering
\caption{Reference definitions of the eight perturbation priors. $w(\vk)$ is the Fourier-space shaping weight applied to white Gaussian noise; all priors are normalized to unit variance after shaping.}\label{tab:noise_defs}
\footnotesize
\begin{tabular}{@{}p{0.24\columnwidth} p{0.32\columnwidth} p{0.34\columnwidth}@{}}
\toprule
Prior & $w(\vk)$ & Intent \\
\midrule
White & $1$ & Baseline; Gaussian prior from DDPM \\
Blue ($\beta{=}0.5,1.0$) & $k^{\beta}$ & Suppress low-$k$; push noise mass to high frequencies \\
Spectrum-matched & $\sqrt{E(k)}$ & Align prior with turbulence energy spectrum \\
Inverse-spectrum & $1/\sqrt{E(k)}$ & Emphasize under-energetic high-$k$ modes \\
Error-weighted & $\sqrt{\mathrm{MSE}_{\mathrm{base}}(k)}$ & Shape by base model's per-$k$ prediction error \\
Von K\'{a}rm\'{a}n & $k^{2}(1+(kL)^{2})^{-17/6}$ & Classical turbulence-model prior \\
Div-free white & Leray-projected white noise & Structural prior with $\nabla\!\cdot\veps{=}0$ \\
\bottomrule
\end{tabular}
\end{table}

\begin{table}[!htbp]
\centering
\caption{Noise type ablation. $\Delta$: change vs.\ white noise. All types normalized to unit variance.}\label{tab:noise}
\footnotesize
\begin{tabular}{@{}lcccc@{}}
\toprule
& \multicolumn{2}{c}{TGV} & \multicolumn{2}{c}{FIT} \\
\cmidrule(lr){2-3} \cmidrule(lr){4-5}
Noise type & RMSE & $\Delta$ & RMSE & $\Delta$ \\
\midrule
White (baseline) & \textbf{0.0371} & --- & 0.0762 & --- \\
Blue ($\beta{=}0.5$) & 0.0382 & +3.0\% & \textbf{0.0754} & $-$1.0\% \\
Blue ($\beta{=}1.0$) & 0.0394 & +6.2\% & 0.0769 & +0.9\% \\
Spectrum-matched & 0.0389 & +4.8\% & 0.0771 & +1.2\% \\
Inverse-spectrum & 0.0391 & +5.4\% & 0.0757 & $-$0.6\% \\
Error-weighted & 0.0387 & +4.3\% & 0.0768 & +0.8\% \\
Von K\'{a}rm\'{a}n & 0.0395 & +6.5\% & 0.0773 & +1.4\% \\
Div-free white & 0.0384 & +3.5\% & 0.0766 & +0.5\% \\
\bottomrule
\end{tabular}
\end{table}

\noindent\textbf{Cross-dataset sensitivity.} The spread of RMSE across the eight priors is one-sided and up to $+6.5\%$ on TGV (every non-white prior is worse than white), while on FIT it is two-sided and narrower, within $[-1.0\%, +1.4\%]$. This reversal is intuitive once considered through the two spectra in \cref{fig:spectra}: TGV's energy is concentrated in a few low-$k$ modes, so any departure from white noise selectively amplifies or damps these dominant modes and is therefore disproportionately visible; FIT's broad inertial cascade averages over a wide range of $k$, and any single shaping function affects many modes weakly. The prior therefore acts as a near-trivial knob on FIT and a slightly sharper one on TGV, but in both cases the effect is substantially smaller than the structural design choices studied in the main paper.

\noindent\textbf{High-pass priors on FIT.} Among the eight, only two priors improve over white on FIT: Blue ($\beta{=}0.5$) at $-1.0\%$ and Inverse-spectrum at $-0.6\%$. Both put additional perturbation mass at higher $k$, which is also the band where FIT's residual error concentrates. The pattern is not monotone in $\beta$ (at stronger high-pass $\beta{=}1.0$, the small gain disappears and turns slightly negative at $+0.9\%$), but it is consistent with the view that perturbing where the base predictor is weakest yields small benefits. On TGV the opposite holds: every non-white prior is worse than white, because there is no meaningful high-$k$ error to shape toward.

The overall variation across priors (within a $7\%$ band on TGV and a $2.4\%$ band on FIT) is substantially smaller than the sensitivity to refinement depth $K$, the sigma range $\sigma_{\max}$, or the schedule choice (all studied in \cref{sec:exp_main}): increasing $\sigma_{\max}$ from $0.01$ to $0.05$ already worsens FIT R1 by $14\%$. Noise prior design is therefore a secondary tuning knob rather than a primary design axis, confirming the claim in \cref{sec:discussion}.

\begin{figure*}[!t]
\centering
\includegraphics[width=\textwidth]{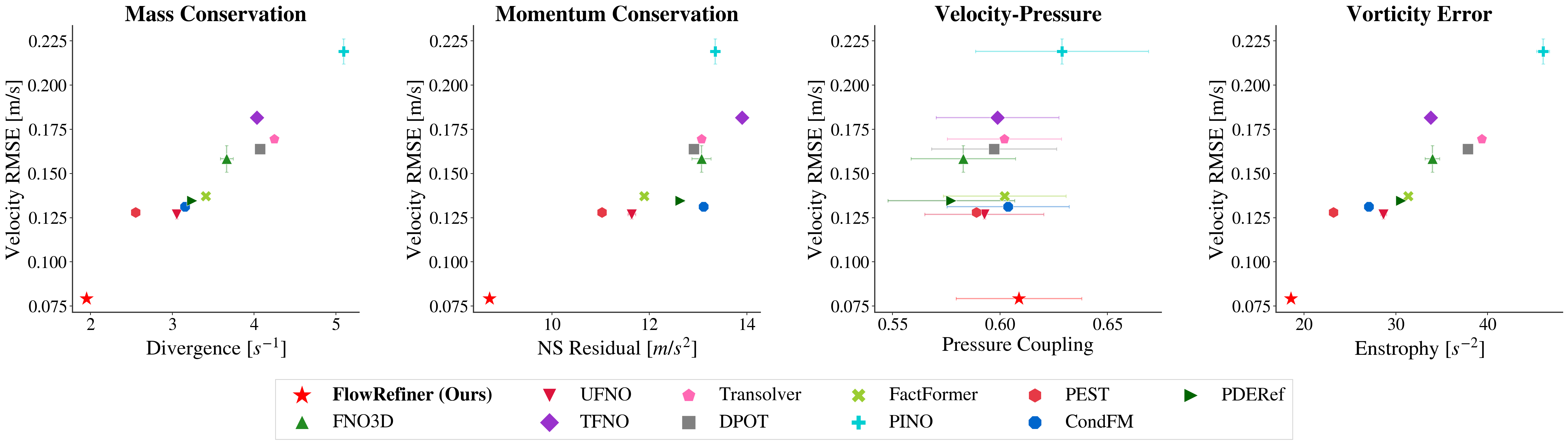}
\caption{Physical consistency analysis on FIT at autoregressive Round~1. Axes as in \cref{fig:physics}.}\label{fig:physics_r1}
\end{figure*}

\begin{figure*}[!t]
\centering
\includegraphics[width=\textwidth]{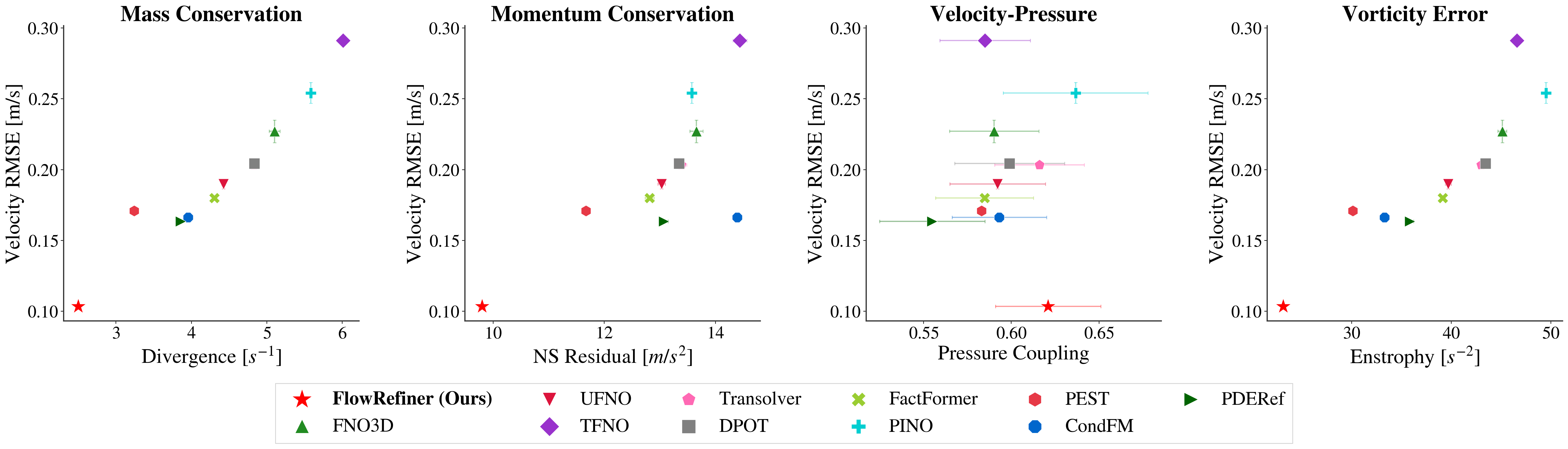}
\caption{Physical consistency analysis on FIT at autoregressive Round~2. Axes as in \cref{fig:physics}.}\label{fig:physics_r2}
\end{figure*}

\section{Physical Consistency at Earlier Rollout Rounds.}\label{app:physics_rounds}

\Cref{fig:physics_r1,fig:physics_r2} show the physical consistency analysis at autoregressive Rounds~1 and~2, complementing the Round~3 results in \cref{fig:physics}.
The same trend holds across all rounds: FlowRefiner achieves the lowest velocity RMSE while maintaining physical residuals comparable to physics-supervised baselines, confirming that the implicit physical consistency observed at Round~3 is not an artifact of late-rollout error accumulation.

\noindent\textbf{Round 1 observations.}
At R1, FlowRefiner (red star) sits in the lower-left of every one of the four panels, with both the lowest velocity RMSE and low-to-moderate physical residuals. The ten baselines shown cluster in the upper-right quadrant with visibly larger velocity error at comparable or worse physics (we omit DiT-Ref here for readability, since its accuracy on FIT is far off the plotted axes). The most informative comparison is against PEST (red circle), which explicitly supervises the Navier--Stokes residual and the divergence constraint during training: PEST is the closest baseline to FlowRefiner along the horizontal axis (physical residual) in every panel, but remains clearly above FlowRefiner along the vertical axis (velocity RMSE). In other words, matching PEST on physics does not require an explicit physics loss; it emerges from refinement on top of a strong base predictor. We also note that the spectral and attention-based operators (FNO3D, TFNO, PINO, Transolver, DPOT, FactFormer) are separated from FlowRefiner on both axes, indicating that the R1 gap is not a velocity-only phenomenon.

\noindent\textbf{Round 2 observations.}
From R1 to R2 the entire scatter cloud drifts upward and to the right, reflecting that both velocity RMSE and every physics metric grow with rollout depth for all methods. However, the \emph{relative} ordering of methods is preserved across rounds: FlowRefiner remains in the lower-left on all four panels, PEST remains the closest competitor on physics but distinctly worse on velocity, and the operator/attention baselines remain separated from both. Comparing the R1, R2, and R3 panels (\cref{fig:physics_r1,fig:physics_r2,fig:physics}), FlowRefiner stays clearly separated from the baseline cluster at every round, with no round where a baseline catches up on either axis.

The implicit physical consistency observed in the main paper (Round 3) is already present at Round 1. This matters because it rules out an alternative explanation in which FlowRefiner's physics advantage would only appear once baselines have accumulated enough rollout error to drift off the physics manifold; instead, the advantage is established from the very first block prediction, and is then sustained rather than generated by the rollout.

\end{document}